\documentclass[a4paper,11pt]{article}
\pdfoutput=1 

\usepackage{jheppub} 

\usepackage{amsmath}
\usepackage{amssymb}
\usepackage{young}
\usepackage[vcentermath]{youngtab}
\usepackage{dsfont}
\usepackage{braket}
\usepackage{comment}
\usepackage{textgreek} 
\usepackage{simplewick}
\usepackage{scalerel}

\allowdisplaybreaks
\usepackage{mathrsfs}
\usepackage[mathscr]{euscript}
\usepackage{tikz}
\usepackage{empheq}
\usepackage[many]{tcolorbox}

\newtcolorbox{empheqboxed}{colback=gray!30, 
 colframe=white,
 width=\textwidth,
 sharpish corners,
 top=-2mm, 
 bottom=0pt
}

\numberwithin{equation}{section}

\def\be{\begin{equation}}
\def\ee{\end{equation}}

\newcommand{\lgamma}{\text{\textgamma}}

\newcommand{\hypgeo}[2]{%
  {\vphantom{F}}_{#1}\kern-\scriptspace F_{#2}%
  }

\title{
String correlators on $\boldsymbol{\text{AdS}_3}$: Four-point functions}

\author[a]{Andrea Dei,}
\author[b]{Lorenz Eberhardt}
\affiliation[a]{Jefferson Physical Laboratory, Harvard University, \\
\hspace*{0.3cm}Cambridge, MA 02138 USA}
\affiliation[b]{School of Natural Sciences, Institute for Advanced Study, \\
\hspace*{0.3cm}Princeton, NJ 08540, USA}
\emailAdd{adei@fas.harvard.edu}
\emailAdd{elorenz@ias.edu}

\abstract{We propose a closed-form formula for genus 0 four-point functions in $\text{AdS}_3$ string theory with pure NS-NS flux including arbitrary amounts of spectral flow. Our formula passes many non-trivial consistency checks and has intriguing connections to Hurwitz theory. This paper is the second in a series with several instalments.}

\begin{document}
\maketitle
\flushbottom

\section{Introduction}

The duality between strings propagating on three-dimensional Anti-de Sitter space ($\text{AdS}_3$) and two-dimensional conformal field theories ($\text{CFT}_2$) is one of the best understood incarnations of the AdS/CFT correspondence \cite{Maldacena:1997re}. Due to the presence of R-R fluxes, the study of higher-dimensional holographic setups is much more difficult from a worldsheet CFT perspective. Strings propagating on $\text{AdS}_3$ with pure NS-NS flux can be studied via the $\text{SL}(2,\mathds{R})$ Wess-Zumino-Witten (WZW) model. This has a long history \cite{Gawedzki:1991yu,Teschner:1997ft,Giveon:1998ns,deBoer:1998gyt,Kutasov:1999xu,Teschner:1999ug,Maldacena:2000hw, Maldacena:2000kv, Maldacena:2001km} and repeated advances throughout the last three decades have raised the hope to completely solve the worldsheet model and compute string theory observables exactly at tree level.

A complete understanding of the spectrum of the $\text{SL}(2,\mathds{R})$ WZW model has not been reached until the need to introduce spectral flow was realised \cite{Balog:1988jb, Petropoulos:1989fc, Hwang:1990aq, Henningson:1991jc, Gawedzki:1991yu, Bars:1995mf, Teschner:1997ft, Evans:1998qu, Giveon:1998ns, deBoer:1998gyt, Teschner:1999ug, Kutasov:1999xu, Giribet:1999ft, Giribet:2000fy, Maldacena:2000hw, Maldacena:2000kv, Maldacena:2001km, Giribet:2001ft}. The presence of spectrally flowed representations --- corresponding to non highest-weight modules of the worldsheet algebra --- provides a rich and intriguing dynamics for $\text{AdS}_3$ strings. At the same time, while the spectrum of the worldsheet theory is under a firm control, the presence of spectral flow has hampered a full understanding of $\text{AdS}_3$ string correlators. See \cite{Giribet:2000fy,Stoyanovsky:2000pg,Fateev,Maldacena:2001km,Giribet:2001ft,Giribet:2005ix,Ribault:2005ms,Minces:2005nb,Ribault:2005wp,Iguri:2007af,Hikida:2007tq,Baron:2008qf,Iguri:2009cf,Giribet:2011xf,Cagnacci:2013ufa,Cagnacci:2015pka,Giribet:2019new,Eberhardt:2019ywk,Hikida:2020kil} for some important results. 

Recently, a further step towards a full solution of the $\text{SL}(2,\mathds{R})$ WZW model has been taken in \cite{Dei:2021xgh}, where we proposed a closed-form formula for three-point functions with an arbitrary amount of spectral flow. As usual in conformal field theory, our formula for the structure constants is just a collection of numbers since the coordinate dependence of three-point functions is fixed by the global conformal symmetry on the worldsheet and in spacetime. Four- and higher-point functions contain much more interesting dynamical information. While in principle they can be accessed by a conformal block expansion, it is presently not known how to formulate such a conformal block expansion that fully incorporates spectral flow in the $x$-basis. For this reason, it is very worth-while to study four-point functions directly without making use of a spectrally flowed conformal block expansion.
 
We exclusively study bosonic strings at genus 0 in this paper. On the worldsheet, this means that we focus on the CFT describing bosonic strings on Euclidean $\text{AdS}_3$ defined by analytic continuation of the $\text{SL}(2,\mathds{R})_k$ WZW model --- see \cite{Maldacena:2001km} for the precise definition of the worldsheet theory. We follow the conventions of our previous paper \cite{Dei:2021xgh}.
This is however not a restriction since the description of superstrings in the RNS-formalism involves the same model at level $k+2$, together with free fermions and at least at genus 0 the two sectors are completely decoupled. Thus we think that most of our results carry over to the superstring case with the replacement $k \to k+2$ in most of the formulae.\footnote{When computing genus zero superstring $n$-point functions via the RNS-formalism, one also needs to insert $n-2$ picture changing operators. This is in general a non-trivial step, even though the necessary technology is well-known in the literature. See for example \cite{Green-Schwarz-Witten,Gaberdiel:2007vu,Dabholkar:2007ey}.}

\begin{figure}
\begin{center}
\includegraphics[trim=7.7cm 10.8cm 10.5cm 10.5cm,clip, width=0.7\textwidth]{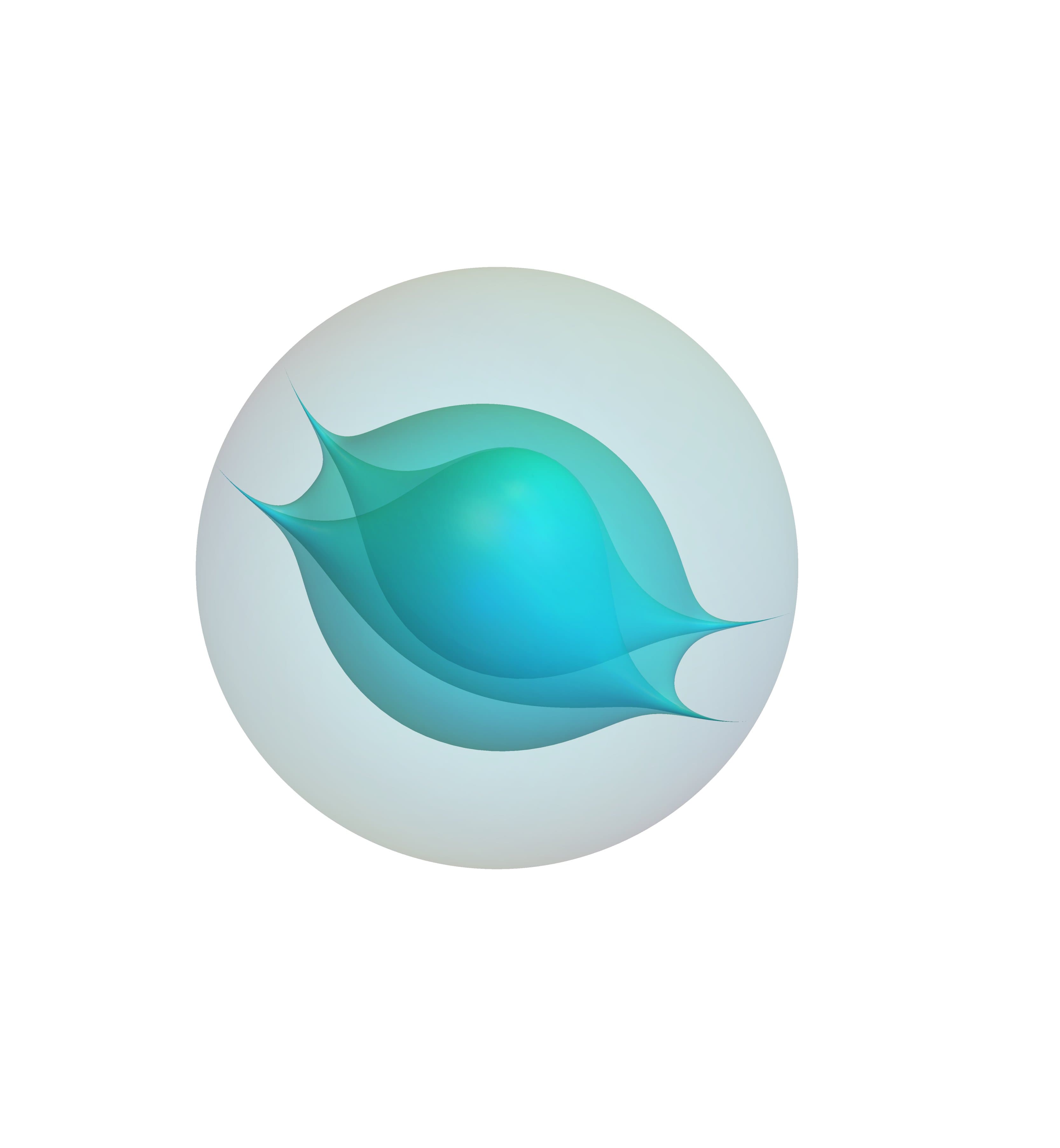}	
\end{center}
\caption{A sketch of the string interaction in $\text{AdS}_3$.}
\label{fig:string interaction}
\end{figure}

We are then motivated to continue our study of spectrally flowed correlators on $\text{AdS}_3$. Let us explain the most important features with the help of Figure \ref{fig:string interaction} --- more precise definitions will be spelled out in the following sections. Euclidean $\text{AdS}_3$ is topologically a three-dimensional ball with the Riemann sphere $\text{S}^2$ as asymptotic boundary. The worldsheet of the four-point string correlator whose vertex operators carry two units of spectral flow is depicted in Figure~\ref{fig:string interaction}. As one can see, there are two sheets of the worldsheet touching each insertion point corresponding to the two units of spectral flow. In total, there are three sheets, ensuring that the topology of the worldsheet is a Riemann sphere. 
 
\medskip

Vertex operators describe the insertion of an asymptotic string state. From the cartoon of Figure~\ref{fig:string interaction}, one directly sees that there are two coordinates associated with the vertex operator: the worldsheet coordinate $z$ and the boundary coordinate $x$, where the worldsheet touches the boundary. Moreover, one can see that there is another integer associated to each vertex operator --- the so-called spectral flow $w \in \mathds{Z}_{\ge 0}$. It corresponds to the number of times the worldsheet is winding around the insertion point in the boundary.\footnote{This is not a topological invariant and from the picture alone it is not clear whether this number remains well-defined in the quantum theory, since it seems that there is geometrically a continuous transition between the different winding numbers. However, this integer will correspond to a certain well-defined number that labels the representation of the symmetry algebra $\mathfrak{sl}(2,\mathds{R})_k\times \mathfrak{sl}(2,\mathds{R})_k$.} 
Beyond these geometric quantities, the vertex operators have also two different $\mathfrak{sl}(2,\mathds{R})$ spins associated to them. In spacetime, they are primary operators of spacetime conformal weight $(h,\bar{h})$ and on the worldsheet they are built on $\mathfrak{sl}(2,\mathds{R})$ representations of spin $j$. While as usual in a unitary CFT $h-\bar{h} \in \mathds{Z}$ and $h,\bar{h} \ge 0$, the $\mathfrak{sl}(2,\mathds{R})$ spin on the worldsheet can fall into either the discrete series with $j \in \mathds{R}$ or the principal continuous series with $j \in \tfrac{1}{2}+i \, \mathds{R}$.\footnote{The case $w=0$ is exceptional, since $j$ and $h$ coincide.}

To summarise, the worldsheet vertex operators take in general the following form
\be 
V_{j,h,\bar{h}}^w(x;z)\ .
\ee
Thus, the objects that we want to study are the correlators
\be 
\left\langle V_{j_1,h_1,\bar{h}_1}^{w_1}(x_1;z_1) \cdots  V_{j_n,h_n,\bar{h}_n}^{w_n}(x_n;z_n) \right \rangle\ .
\ee
In this paper we take a step towards the full understanding of such correlators. For genus 0 four-point functions, we propose a map that takes an unflowed correlation function and transforms it into a flowed correlation function with the desired spectral flow indices. 
Notice that this is not just a rewriting of the problem, since in the unflowed sector (i.e.~the case where all vertex operators satisfy $w_i=0$) these correlators are very well-known and have been thoroughly studied in the literature, see e.g.~\cite{Teschner:1997ft,Teschner:1999ug,Giribet:2000fy,Giribet:2001ft,Teschner:2001gi,Maldacena:2001km,Ribault:2005wp,Hikida:2007tq}. 

Our approach is almost entirely based on symmetry. We use all available symmetries on the worldsheet to constrain correlators as much as possible. These constraints come in the form of global and local Ward identities \cite{Eberhardt:2019ywk}, Knizhnik-Zamolodchikov \cite{Knizhnik:1984nr} equations and null vector equations. Most of these are much more involved than for the unflowed sector and not known in closed-form for generic choice of the spectral flow parameters. Nevertheless, by explicitly solving these constraints we end up with a relatively simple closed-form formula. For the four point function, it is given in the form of an integral transform relating the unflowed sector to the flowed sector,
\begin{subequations}
\begin{tcolorbox}[ams align]
&\hspace{-30pt}\left\langle V_{j_1,h_1, \bar h_1}^{w_1}(0;0) V_{j_2,h_2, \bar h_2}^{w_2}(1;1) V_{j_3,h_3, \bar h_3}^{w_3}(\infty;\infty) V_{j_4,h_4, \bar h_4}^{w_4}(x;z) \right \rangle \nonumber\\
& \hspace{-13pt} = \int \prod_{i=1}^4 \mathrm{d}^2 y_i \ y_i^{\frac{kw_i}{2}+j_i-h_i-1}\bar{y_i}^{\frac{kw_i}{2}+j_i-\bar{h_i}-1} |X_\emptyset|^{2(j_1+j_2+j_3+j_4-k)} \nonumber  \\
& \quad \quad\times |X_{12}|^{2(-j_1-j_2+j_3-j_4)} \, |X_{13}|^{2(-j_1+j_2-j_3+j_4)} |X_{23}|^{2(j_1-j_2-j_3+j_4)} |X_{34}|^{-4j_4} \nonumber \\
& \quad \quad\times\left\langle V_{j_1}^{0}(0;0) V_{j_2}^{0}(1;1) V_{j_3}^{0}(\infty;\infty) V_{j_4}^{0}\left(\frac{X_{23}X_{14}}{X_{12}X_{34}};z\right) \right \rangle\, 
\label{eq:main conjecture even}
\end{tcolorbox}
\noindent for $\sum_i w_i \in 2 \mathds{Z}$ and
\begin{tcolorbox}[ams align]
&\hspace{-21pt} \left\langle V_{j_1,h_1, \bar h_1}^{w_1}(0;0) V_{j_2,h_2, \bar h_2}^{w_2}(1;1) V_{j_3,h_3, \bar h_3}^{w_3}(\infty;\infty) V_{j_4,h_4, \bar h_4}^{w_4}(x;z) \right \rangle \nonumber\\
&\hspace{-5pt}=\mathcal{N}(j_3) \int \prod_{i=1}^4 \mathrm{d}^2 y_i \ y_i^{\frac{kw_i}{2}+j_i-h_i-1}\bar{y_i}^{\frac{kw_i}{2}+j_i-\bar{h_i}-1} |X_{123}|^{2(\frac{k}{2}-j_1-j_2-j_3-j_4)} \nonumber \\
& \quad \quad \quad\times |X_{1}|^{2(-j_1+j_2+j_3+j_4-\frac{k}{2})} \, |X_{2}|^{2(j_1-j_2+j_3+j_4-\frac{k}{2})} |X_3|^{2(j_1+j_2-j_3+j_4-\frac{k}{2})} |X_{4}|^{-4 j_4} \nonumber \\
&\quad \quad \quad\times\left\langle V_{j_1}^{0}(0;0) V_{j_2}^{0}(1;1) V_{\frac{k}{2}-j_3}^{0}(\infty;\infty) V_{j_4}^{0}\left(\frac{X_{2}X_{134}}{X_{123}X_{4}};z\right) \right \rangle\, \ , 
\label{eq:main conjecture odd}
\end{tcolorbox}
\label{eq:main conjecture}
\end{subequations}
\noindent for $\sum_i w_i \in 2 \mathds{Z}+1$. 

Let us point the reader to the relevant definitions that enter these formulae. First of all, the presence of the $y$-integral transform was explained in detail in \cite{Dei:2021xgh} and we recall the main definition in \eqref{eq:definition y basis}. 
The quantities $X_I$ are polynomials of degree 1 in each $y_i$ for $i \in I \subset \{1,2,3,4\}$. They are defined in eq.~\eqref{eq:def XI}. In that definition the quantities $P_{\boldsymbol{w}}(x;z)$ enter; they are the (suitably normalised) polynomials that carve out the Hurwitz space of ramified covers, see eqs.~\eqref{eq:def Ptilde poly} and \eqref{eq:def P poly}. Finally, the prefactor $\mathcal{N}(j)$ in \eqref{eq:main conjecture odd} is constructed out of the normalisation of the two-point function, see eq.~\eqref{Nj}. The positions of the first three vertex operators can be chosen to be arbitrary when making use of global Ward identities, see Section~\ref{sec:global ward id}.

The correlator vanishes identically for $\sum_i w_i \le 2\max_i (w_i)-3$ \cite{Maldacena:2001km,Eberhardt:2019ywk}, which is reflected in the vanishing of the quantities $X_I$. For the edge case $\sum_i w_i=2\max_i(w_i)-2$, the expression is still valid, provided it is interpreted correctly. In our expression, both $X_\emptyset$ and the unflowed correlator that enters the right-hand side are singular. Their product is still well-behaved and can be defined by a limiting procedure as explained in Section~\ref{sec:boundary w}. We should also mention that this formula as written is valid for continuous representations $j_i \in \frac{1}{2}+i\, \mathds{R}$. For discrete representations there is a small modification in the integration contours as discussed in Section~\ref{subsec:review}.

Eq.~\eqref{eq:main conjecture} essentially encodes the geometry of the problem. One can already guess from Figure~\ref{fig:string interaction} that the relevant geometric notion that enters the stage is a holomorphic branched covering map. In fact, it was argued in \cite{Maldacena:2001km} that configurations in the worldsheet field space for which the worldsheet covers the boundary holomorphically are somewhat analogous to worldsheet instanton sectors, although strictly speaking these are not different topological sectors.

\medskip 

We will show that our formula \eqref{eq:main conjecture} satisfies all the constraints that follow from symmetry, namely
\begin{itemize}
\item Global Ward identities (both in $x$- and $z$-space),
\item Local Ward identities,
\item Knizhnik-Zamolodchikov equation,
\item Null vector decoupling (when applicable). 
\end{itemize}
These constraints fix the result essentially uniquely up to a constant (i.e.\ independent of the coordinates $x_i$, $y_i$ and $z_i$) prefactor. 
We perform several consistency checks that strongly motivate the prefactor. These are
\begin{itemize}
\item Internal consistency,
\item Contact with previous results in the literature for special cases,
\item Reduction to our previous conjecture about the three-point function \cite{Dei:2021xgh},
\item Reflection symmetry for continuous representations, 
\item Bosonic exchange statistics.
\end{itemize}
Some of these seem obvious (especially the last points), but it turns out that they impose lots of non-trivial constraints on our proposal. 

\medskip

This paper is computationally quite heavy. Because of the way the various constraints are formulated on the correlators we can only evaluate them for fixed values of spectral flow. So our strategy is to systematically solve them for a large number of choices of spectral flows (we have analyzed 721 cases) and try to guess the general formula from these examples. While we have managed to find a simple formula that works for all cases it might be hard for the reader to understand where certain formulae come from. For this reason we have included a \texttt{Mathematica} notebook that performs many of the computations that are described in this paper. We have tried to keep the code general and flexible in the hope that it will be useful to some of the readers.

The solution \eqref{eq:main conjecture} depends very much on the existence of polynomials $P_{\boldsymbol w}(x;z)$ whose zero-locus is the so-called \emph{Hurwitz-space} of branched covers from the sphere to itself with ramification indices $\boldsymbol{w}=(w_1,w_2,w_3,w_4)$ at four points. The precise definition is given in eq.~\eqref{eq:def P poly}. These polynomials control the complete flowed structure of the correlator in terms of the unflowed correlator. Consistency of our proposal \eqref{eq:main conjecture} implies various identities among these polynomials, which we collect in Appendix~\ref{app:Pw-identities}. We have tested them in multiple examples in the ancillary {\tt Mathematica} notebook and we have a certain degree of confidence in their correctness. To the best of our knowledge these identities are not known in the mathematics literature and it would be interesting to deduce them from a more rigorous point of view in the context of Hurwitz theory. 

\medskip

The paper is organised as follows. After briefly recalling the results of \cite{Dei:2021xgh}, in Section~\ref{sec:constraints} we describe the various symmetry constraints that spectrally flowed correlators must obey. As far as we are aware, the KZ-equations and the null vector equations in the spectrally flowed sector have not appeared before in the literature.\footnote{See however \cite{Ribault:2005ms} for an analysis of the spectrally flowed KZ-equation in the $m$-basis. This corresponds to a degenerate case of our KZ-equation where several vertex operators collide in spacetime. } In Section~\ref{sec:4ptf} we then move on to the object of ultimate interest --- the four-point function. It depends non-trivially on \emph{two} crossratios (one on the worldsheet and one in spacetime). We fix the dependence on the coordinates fully in terms of the corresponding unflowed correlator. We analyse further properties and carry out various consistency checks on our proposal in Section~\ref{sec:checks}. In Section~\ref{sec:conclusions} we comment on various points deserving a better understanding and suggest future directions of research. Various appendices complement the discussion of some technical points we encounter throughout the text. 

\section{Constraints}
\label{sec:constraints}

\subsection{A short review of the three-point function} 
\label{subsec:review}
We will pick the discussion up more or less where we ended it in \cite{Dei:2021xgh}. To make the paper self-contained, we briefly recall the result that we conjectured there for the three-point functions.

\paragraph{Definition of the vertex operators. } We want to compute correlators of the spectrally flowed affine primary vertex operators
\be 
V_{j,h,\bar{h}}^w(x,\bar{x};z,\bar{z})\ .
\label{vertex-op}
\ee
Our definition coincides with that of most of the literature, for example of \cite{Maldacena:2001km}. For the details of how these operators can be constructed we refer to \cite{Dei:2021xgh}. Here, $j$ is the $\mathfrak{sl}(2,\mathds{R})_k$ spin of the vertex operator on the worldsheet. It can take either values in $\frac{1}{2}<j<\frac{k-1}{2}$, corresponding to discrete representations, or values in $\frac{1}{2}+i \mathds{R}$, corresponding to continuous representations. The non-negative integer $w$ parametrises the amount of spectral flow. In particular, $w=0$ labels the unflowed sector. $(h,\bar{h})$ are the conformal weights of the operator in \emph{spacetime} (they are denoted by $(J,\bar{J})$ in \cite{Maldacena:2001km}). Finally, $x$ and $z$ are the insertion points of the operator in spacetime and on the worldsheet respectively. In the unflowed sector for $w=0$ the labels $j$ and $h=\bar{h}$ coincide and we will hence omit the labels $h$ and $\bar{h}$. We will also frequently omit the right-moving labels $\bar{h}$, $\bar{x}$ and $\bar{z}$ in the flowed sector.

\paragraph{$\boldsymbol{y}$-basis.} In \cite{Dei:2021xgh} we introduced an integral transform that trades the conformal weights $(h,\bar{h})$ for a complex variable $y$ as follows:
\be 
V_{j,h,\bar{h}}^w(x;z)=\int_{\mathfrak{C}} \mathrm{d}^2 y\ y^{\frac{kw}{2}+j-h-1}\bar{y}^{\frac{kw}{2}+j-\bar{h}-1} V_{j}^w(x;y;z)\ .
\label{eq:definition y basis}
\ee 
The contour of integration $\mathfrak{C}$ depends on the type of representation. For a continuous representation one integrates over the whole complex plane, whereas for a discrete representation one takes a contour integral around $0$ (in the case of a lowest weight representation $\mathcal{D}^+$) or $\infty$ (in the case of a highest weight representation $\mathcal{D}^-$) for both $y$ and $\bar{y}$. Most of our results look simple when expressed in the $y$-basis, but computing the integral over $y$-space --- which is necessary in order to transform the expression back to the $h$-variables --- is usually complicated.

\paragraph{Reflection symmetry.} Let us mention one further aspect of the $y$-basis. The continuous representations with spins $j$ and $1-j$ are equivalent. Similarly to what happens for unflowed vertex operators in the $x$-basis \cite{Teschner:1997ft},
\be 
V^0_{1-j}(x;z) = R_{1-j} \frac{(1-2j)}{\pi} \int \mathrm{d}^2 x' \ |x-x'|^{4j-4} \,  V^0_{j}(x';z) \ . 
\label{eq:Teschner-reflection}
\ee
this equivalence is expressed for flowed vertex operators in the $y$-basis via a simple integral transform
\be 
V^w_{1-j}(x;y;z) = R_{1-j} \frac{(1-2j)}{\pi} \int \mathrm{d}^2 y' \ |y-y'|^{4j-4} \,  V^w_{j}(x;y';z) \ ,
\label{eq:y-basis-reflection}
\ee
where $R_j$ is the reflection coefficient. It is related to the normalisation $B(j)$ of the two-point function via
\be 
R_j=\frac{\pi B(j)}{2j-1} \label{eq:reflection coefficient}
\ee
and satisfies $R_jR_{1-j}=1$. 

\paragraph{Three-point functions.} Our main conjecture in \cite{Dei:2021xgh} was a closed-form formula for the three-point functions of these operators. To simplify, we used global Ward identities in both $x$- and $z$-space to put the operators at $x_1=z_1=0$, $x_2=z_2=1$ and $x_3=z_3=\infty$. Our result was then
\begin{align}
&\left\langle V^{w_1}_{j_1}(0; y_1; 0) \,  V^{w_2}_{j_2}(1; y_2; 1) \, V^{w_3}_{j_3}(\infty; y_3; \infty) \right\rangle \nonumber\\
&\quad=\begin{cases} D(j_1,j_2,j_3)|X_\emptyset|^{2\sum_l j_l-2k}\displaystyle\prod_{i<\ell}^3  |X_{i \ell}|^{2\sum_l j_l-4j_i-4j_\ell} \  , & \sum_i w_i \in 2\mathds{Z}\ ,\\
\mathcal{N}(j_1) D(\tfrac{k}{2}-j_1,j_2,j_3) |X_{123}|^{k-2\sum_l j_l} \displaystyle\prod_{i=1}^3 |X_i|^{2\sum_l j_l-4j_i-k} \ ,  & \sum_i w_i \in 2\mathds{Z}+1\ .
\end{cases} \label{eq:three-point function}
\end{align}
Here, $D(j_1,j_2,j_3)$ are the structure constants of the unflowed three-point functions. They are obtained from those of the $H_3^+$ model by analytic continuation and were determined by Teschner in \cite{Teschner:1997ft}, see \cite{Maldacena:2001km} for our conventions. The quantity $\mathcal{N}(j)=\sqrt{B(j)\big/ B\left(\frac{k}{2}-j\right)}$ is determined through the normalisation of the unflowed two-point function. The quantities $X_I$ for $I \subset \{1,2,3\}$ encode the dependence on the spectral flow. They are linear functions in $y_i$ for $i \in I$ and we gave a very explicit formula for the coefficients in \cite{Dei:2021xgh}. For example, for $\boldsymbol{w}=(6,7,9)$,
\be 
X_{12}=116424 - 2520 y_1 - 5292 y_2 - 
 126 y_1y_2\ ,
\ee
so everything is very explicit. It follows from the explicit formula for $X_I$ that the three-point functions are non-vanishing whenever the spectral flow indices satisfy
\be 
\sum_{i=1}^3 w_i \ge 2\max_i (w_i)-1\ .
\ee
This is a manifestation of the fusion rules of the model and agrees with the analysis of \cite{Maldacena:2001km} and \cite{Eberhardt:2019ywk}. Our main objective in this paper will be to demonstrate that the formula \eqref{eq:three-point function} admits a natural generalisation to the case of four-point functions.

\subsection{Covering maps}
\label{subsec:covering maps}
One of the main players in this paper will be branched covering maps, since it will turn out that the generalisation of the formula \eqref{eq:three-point function} to four-point functions heavily relies on them. We will collect here all the relevant properties of such covering maps. For the purpose of this paper, we will only consider branched coverings from the sphere to itself, $\gamma: \text{S}^2 \longrightarrow \text{S}^2$. When we want to emphasise the complex structure on $\text{S}^2$ in the following, we write $\text{S}^2=\mathbb{CP}^1$. $\gamma$ is a branched covering map, which means that it is locally biholomorphic, except at finitely many points $z_1, \dots, z_n$, where $\gamma$ has ramification indices $w_i\in \mathds{Z}_{\ge 0}$. In other words, the only critical points of $\gamma$ are $z_i$, where it has the Taylor expansion
\be 
\gamma(\zeta)=x_i+a_i(\zeta-z_i)^{w_i}+\mathcal{O}\big((\zeta-z_i)^{w_i+1}\big)\ .
\label{eq:covering map expansion}
\ee
In the mathematics literature one usually assumes that $w_i \ge 2$ so that the map is really ramified there. But to ensure uniformity of our discussion it is also useful to allow for $w_i=0$ and $w_i=1$.
Of course when $w_i=1$, the point is unramified and the only requirement is that $z_i$ is mapped to $x_i$. For $w_i=0$ there is no condition on the map $\gamma$ at the point $z_i$ (and hence we could delete $x_i$ and $z_i$ from the list of points). 

In this paper we will only consider the case of four ramification points. In this case, we can set $x_1=0$, $x_2=1$, $x_3=\infty$, $x_4=x$ and the same for the $z_i$'s.\footnote{Around $\infty$, the relevant Laurent expansion takes the form
\be 
\gamma(\zeta)= \frac{(-1)^{w_3+1}}{a_3} \zeta^{w_3}+\mathcal{O}(\zeta^{w_3-1})\ .
\ee 
}
Given $(w_1,w_2,w_3,w_4)$ and $x$, there are in general only finitely many values of $z$ such that a covering map exists.

The simplest case is given by $w_1=w_2=w_3=w_4=1$, where the covering map has no critical point and is hence a M\"obius transformation. Since it has to map $0$ to $0$, $1$ to $1$ and $\infty$ to $\infty$, the only possible choice is $\gamma(\zeta)=\zeta$. This is only possible provided that $z=x$ and hence only in this case a covering map exists. In general, the existence of a covering map imposes a relation between $x$ and $z$. Since all the conditions we imposed are algebraic, this relation can be written as
\be 
\tilde{P}_{\boldsymbol{w}}(x;z)=0\ ,
\ee
where $\tilde{P}_{\boldsymbol{w}}(x;z)$ is a polynomial in $x$ and $z$. Here and in the following, we lighten the notation by writing $\boldsymbol{w}=(w_1,w_2,w_3,w_4)$. This polynomial defines a subvariety of $\mathbb{CP}^1 \times \mathbb{CP}^1$, which in the mathematics literature is known as the Hurwitz space.\footnote{Strictly speaking the Hurwitz space should be defined as the corresponding subspace of $\mathbb{CP}^1\setminus\{0,1,\infty\} \times \mathbb{CP}^1\setminus\{0,1,\infty\} $ and should be subsequently compactified. This issue will not play a role for the four-point function.} It is a known result that the Hurwitz space is connected and irreducible and hence $\tilde{P}_{\boldsymbol{w}}(x;z)$ is an irreducible polynomial \cite{Liu}. In general, we can write
\be 
\tilde{P}_{\boldsymbol{w}}(x;z)\equiv \prod_{\gamma^{-1}}\left(z-\gamma^{-1}(x)\right)\ , \label{eq:def Ptilde poly}
\ee
where the product runs over all possible preimages of all possible covering maps. The number of terms in the product is known as the \emph{Hurwitz number} and takes in this case the form \cite{Liu}
\be 
H_{\boldsymbol{w}}=\frac{1}{2}\min_{i=1,2,3,4}\Biggl(w_i\Biggl(\sum_j w_j-2w_i\Biggr)\Biggr)\ .
\label{Hurwitz-number}
\ee
This is the order of the polynomial $\tilde{P}_{\boldsymbol{w}}$ in $z$. The order of the polynomial in $x$ is given by\footnote{This can be seen from the symmetric product orbifold CFT, where this number corresponds to the number of channels in the four-point function of four twist-operators. This was discussed in \cite{Dei:2019iym}.} 
\be 
\frac{1}{2} \left(\min(w_1+w_2,w_3+w_4)-\max(|w_1-w_2|,|w_3-w_4|)\right)\ .
\ee
We should note that integrality and non-negativity of $H_{\boldsymbol{w}}$ imposes constraints on $\boldsymbol{w}$. In general a covering map only exists if
\be 
H_{\boldsymbol{w}} \in \mathds{Z}_{\ge 1} \qquad \Longleftrightarrow\qquad \sum_{i=1}^4 w_i \in 2\mathds{Z}_{\ge 1}\ \ \text{and}\ \ \sum_{i=1}^4 w_i >2 \max_{i} w_i\ .
\label{Hurwitz<->coverimg-map}	
\ee
In case $H_{\boldsymbol{w}}=0$, we define $\tilde P_{\boldsymbol{w}}(x;z)=1$, since the product in its definition \eqref{eq:def Ptilde poly} is empty. In case $H_{\boldsymbol{w}}<0$ (or if $w_i<0$ for some $i$), we define $\tilde P_{\boldsymbol{w}}(x;z)=0$.

For illustration, we list in the following a few examples of such polynomials. The definition \eqref{eq:def Ptilde poly} is actually very impracticable to compute these polynomial. In practice, we computed them using an algorithm inspired by \cite{Pakman:2009zz}, which is described in Appendix~\ref{app:algorithm} and implemented in the ancillary {\tt Mathematica} notebook. We have e.g.
\begin{subequations}
\begin{align}
\tilde{P}_{(1,1,1,1)}(x;z)&=z-x \, \\
\tilde{P}_{(2,1,2,1)}(x;z)&=z^2-x \, \\
\tilde{P}_{(2,2,2,2)}(x;z)&=z^4 -4 x z^3+6 x z^2-4 x z+x^2\ , \\
\tilde{P}_{(2,3,4,5)}(x;z)&=z^{10}-30 x z^8+120 x z^7-210 x z^6+168 x z^5-50 x z^4\nonumber\\
&\hspace{30pt} +50 x^2 z^4 -120 x^2 z^3 +105 x^2 z^2-40 x^2 z+6 x^2\ .
\end{align}
\end{subequations}
For later convenience, it will be useful to include prefactors in the definition of the polynomial. These prefactors do not change the interpretation of $P_{\boldsymbol{w}}(x;z)$ as the polynomial defining the Hurwitz space. We define
\begin{multline} 
P_{\boldsymbol{w}}(x;z)\equiv f(\boldsymbol{w}) \, (1-x)^{\frac{1}{2} s\left(-w_1+w_2-w_3+w_4)\right)}(1-z)^{\frac{1}{4} s\left(\left(w_1+w_2-w_3-w_4\right) \left(w_1-w_2-w_3+w_4\right)\right)-\frac{1}{2}w_2w_4}   \\
   \times  x^{\frac{1}{2}s \left(w_1-w_2-w_3+w_4\right)}
 z^{\frac{1}{4}s\left( \left(w_1+w_2-w_3-w_4\right)
   \left(-w_1+w_2-w_3+w_4\right)\right)-\frac{1}{2}w_1w_4}\tilde{P}_{\boldsymbol{w}}(x;z)\ , 
\label{eq:def P poly}
\end{multline}
where $f(\boldsymbol{w})$ is a function of $w_1, \dots, w_4$. The reader can find its explicit form in Appendix~\ref{app:P identities}. Here, we used the function 
\be 
s(\alpha)=\begin{cases}
\alpha \ , \qquad \alpha>0 \ , \\
0 \ , \qquad \alpha\le 0 \ .
\label{s-function}
\end{cases}
\ee
The zero locus of these polynomials coincides with the one of $\tilde{P}_{\boldsymbol{w}}(x;z)$, since we don't allow collisions of $x$ or $z$ with $0$, $1$ or $\infty$. 
The polynomials $P_{\boldsymbol{w}}(x;z)$ satisfy many surprising identities that are relevant for our purposes. We collect them in Appendix~\ref{app:P identities}. We checked many examples in {\tt Mathematica} and have a degree of confidence in their correctness. We do not understand these identities from first principles and a proof would probably enhance our understanding of the story.

Let us remark at this point that upon putting $w_4=0$, one can check that $P_{\boldsymbol{w}}(x;z)$ becomes independent of $x$ and $z$ and hence is a simple number. This is the generalisation of the quantity $P_{\boldsymbol{w}}$ that appeared also in the solution for the three-point functions in \cite{Dei:2021xgh}.

\subsection{The unflowed four-point function}
\label{subsec:unflowed four point function}
Correlators of unflowed vertex operators have been studied thoroughly in the literature. Let us briefly mention the main results and set up our conventions. Two-point functions read \cite{Teschner:1997ft,Fateev,Teschner:1999ug}
\begin{multline}
\left \langle V^0_{j_1}(x_1;z_1) V^0_{j_2}(x_2;z_2) \right \rangle\\
=\frac{1}{|z_1-z_2|^{4 \Delta(j_1)}} \left(\delta^2(x_1-x_2) \delta(j_1+j_2-1)+\frac{B(j_1)}{|x_1-x_2|^{4j_1}} \delta(j_1-j_2) \right)\ , 
\label{eq:unflowed two-point}
\end{multline}
see e.g.~\cite{Dei:2021xgh} for the explicit definition of $B(j)$ and $\Delta(j)=-\frac{j(j-1)}{k-2}$. Let us also define
\be
\mathcal N(j) \equiv \sqrt{\frac{B(j)}{B\left(\frac{k}{2}-j\right)}}=\frac{\nu^{\frac{k}{2}-2j}}{\lgamma\left(\frac{2j-1}{k-2}\right)} \ , 
\label{Nj} 
\ee
which in the following will enter various formulae. Three-point functions are explicitly known \cite{Teschner:1997ft,Fateev,Teschner:1999ug} and we will denote them by 
\be 
D(j_1,j_2,j_3)=\left \langle V^0_{j_1}(0;0)V^0_{j_2}(1;1)V^0_{j_3}(\infty;\infty) \right \rangle\ . \label{eq:definition D}
\ee
Since the generator of translations on the worldsheet can be written as a bilinear in the $\mathfrak{sl}(2,\mathds R)$ currents \cite{Sugawara:1967rw}, 
\begin{align} 
\partial_z V_{j}^0(0;z) & = L_{-1} V_{j}^0(0;z) =  \nonumber \\
& = \frac{1}{k-2} \Bigl(-(J^3 J^3)_{-1} + \tfrac{1}{2} (J^+ J^-)_{-1} + \tfrac{1}{2} (J^- J^+)_{-1} \Bigr) \, V_{j}^0(0;z) \ , 
\label{sugawara}
\end{align}
a partial differential equation for correlators, known as Knizhnik-Zamolodchikov (KZ) equation, can be derived \cite{Knizhnik:1984nr,Frenkel:1991gx,Teschner:1999ug}.  For four-point functions involving only unflowed vertex operators, the KZ equation reads \cite{Teschner:1999ug}\footnote{With respect to the conventions adopted in \cite{Teschner:1999ug}, we have $j_1^{\text{T}}=-j_1$, $j_2^{\text{T}}=-j_4$, $j_3^{\text{T}}=-j_2$ and $j_4^{\text{T}}=-j_3$.}
\begin{multline}
\partial_z \left\langle V^0_{j_1}(0;0) V^0_{j_2}(1;1) V^0_{j_3}(\infty;\infty) V^0_{j_4}(x;z) \right \rangle \\
= \frac{1}{k-2} \Bigl(\frac{\mathcal{P}}{z} + \frac{\mathcal Q}{z-1} \Bigr) \left\langle V^0_{j_1}(0;0) V^0_{j_2}(1;1) V^0_{j_3}(\infty;\infty) V^0_{j_4}(x;z) \right \rangle  \ , 
\label{eq:unflowed-KZ}
\end{multline}
where $\mathcal{P}$ and $\mathcal{Q}$ are differential operators in $x$,
\begin{subequations}
\begin{align}
\mathcal P& = x^{2}(x-1) \partial_x^2-\left((\kappa-1) x^{2}+2j_1x-2 j_4 x(x-1)\right) \partial_x  -2 \kappa j_4 x-2 j_1j_4\ ,  \\
\mathcal Q& =-(1-x)^{2} x \partial_x^2  +(1-x)\left((\kappa-1)(1-x)+2 j_2+2j_4 x\right) \partial_x -2 \kappa j_4(1-x)-2 j_2j_4\ , 
\end{align}
\end{subequations}
and 
\be
\kappa\equiv j_{3}-j_{1}-j_{2}-j_{4} \ . 
\ee 
Beyond the singularities one usually encounters when two insertion points collide on the worldsheet or in spacetime (i.e. when $z=0$, $z=1$, $z= \infty$, $x=0$, $x=1$ or $x= \infty$), 
the unflowed correlator features an additional singularity when $z=x$ \cite{Maldacena:2001km}. This can be seen by substituting the ansatz 
\be 
\left\langle V^0_{j_1}(0;0) V^0_{j_2}(1;1) V^0_{j_3}(\infty;\infty) V^0_{j_4}(x;z) \right \rangle \sim |z-x|^{2 \delta}
\label{unflowed-singular}
\ee
into \eqref{eq:unflowed-KZ}. The first order in $(z-x)$ implies \cite{Maldacena:2001km}
\be 
\delta = 0 \quad \text{or} \quad \delta = k - j_1-j_2-j_3-j_4 \ . 
\label{delta-exp-sing-sol}
\ee 
Both solutions appear in the conformal block expansion and are in general necessary to yield a single-valued four-point function.

Fusion rules and conformal blocks of unflowed correlators are also known \cite{Teschner:1997ft,Teschner:1999ug} and crossing symmetry has been proved in \cite{Teschner:2001gi}.
In the following we will make use of the identity
\begin{multline}
\mathcal N(j_1) \left\langle V^0_{\frac{k}{2}-j_1}(0;0) V^0_{j_2}(1;1) V^0_{j_3}(\infty;\infty) V^0_{j_4}(x;z) \right\rangle \\
= \mathcal N(j_3) |x|^{-4j_4} |z|^{2j_4} \left\langle V^0_{j_1}(0;0) V^0_{j_2}(1;1) V^0_{\frac{k}{2}-j_3}(\infty;\infty) V^0_{j_4}\Bigl(\frac{z}{x};z \Bigr) \right\rangle \ . 
\label{k/2-id}
\end{multline}
Eq.~\eqref{k/2-id} has been proved in \cite{Parnachev:2001gw}. Since various ingredients will be useful in the following, we give an alternative derivation in Appendix~\ref{app:identity}. 

\subsection{Global Ward identities}
\label{sec:global ward id}

Let us now explain the various constraints four-point functions satisfy. We start with the global Ward identities. They allow us to put $x_1=z_1=0$, $x_2=z_2=1$ and $x_3=z_3=\infty$ in our formulae. We discussed them in the $y$-space in \cite{Dei:2021xgh}. One can always reinstate the dependence on all four positions in $x$- and $z$-space by using the solution to the global Ward identities for four-point functions which read
\begin{align}
&\langle V_{j_1}^{w_1}(x_1;y_1;z_1)V_{j_2}^{w_2}(x_2;y_2;z_2)V_{j_3}^{w_3}(x_3;y_3;z_3)V_{j_4}^{w_4}(x_4;y_4;z_4) \rangle \nonumber\\
& =|x_{21}|^{2(-h_1^0-h_2^0+h_3^0-h_4^0)} \, |x_{31}|^{2(-h_1^0+h_2^0-h_3^0+h_4^0)} \, |x_{32}|^{2(h_1^0-h_2^0-h_3^0+h_4^0)} \, |x_{34}|^{-4 h_4^0}\nonumber\\
 &\qquad\!\times  |z_{21}|^{2(-\Delta _1^0-\Delta _2^0+\Delta _3^0-\Delta _4^0)} \, |z_{31}|^{2(-\Delta _1^0+\Delta _2^0-\Delta _3^0+\Delta _4^0)}  \, |z_{32}|^{2(\Delta_1^0-\Delta _2^0-\Delta _3^0+\Delta _4^0)} \,  |z_{34}|^{-4 \Delta _4^0}\nonumber\\
   &\qquad\!\times \Bigg \langle V_{j_1}^{w_1} \left(0;\frac{y_1 \, x_{32} \, z_{21}^{w_1} \, z_{31}^{w_1} }{x_{21} \, x_{31} \, z_{32}^{w_1}};0\right) V_{j_2}^{w_2} \left(1;\frac{y_2 \, x_{31} \, z_{21}^{w_2} \, z_{32}^{w_2}}{x_{21} \, x_{32} \, z_{31}^{w_2}};1\right) V_{j_3}^{w_3}\left(\infty;\frac{y_3 \, x_{21} \, z_{31}^{w_3} \, z_{32}^{w_3}}{x_{31} \, x_{32} \, z_{21}^{w_3}};\infty\right)\nonumber\\
   &\qquad\qquad\qquad\qquad V_{j_4}^{w_4} \left(\frac{x_{32} \, x_{14}}{x_{12} \, x_{34}};\frac{y_4 \, x_{31} \, x_{32} \, z_{21}^{w_4} \,  z_{34}^{2 w_4}}{x_{21} \, x_{34}^2 \, z_{31}^{w_4} \,  z_{32}^{w_4}}; \frac{z_{32} \, z_{14}}{z_{12} \, z_{34}}\right)\Bigg\rangle\ ,\label{eq:global Ward identities solution 4pt function}
\end{align}
where
\be 
h^0\equiv j+\frac{k w}{2}\ , \qquad \Delta^0 \equiv -\frac{j(j-1)}{k-2}-w j-\frac{k w^2}{4}\ . \label{eq:h0 Delta0}
\ee
The prefactors are analogous to those we are used to in CFT. However, since vertex operators in the $y$-basis don't have a definite conformal weight on the worldsheet or in spacetime also the $y$-coordinate transforms non-trivially under conformal transformations.

\subsection{Local Ward identities}

Local Ward identities impose further constraints on correlators. They are a consequence of the \emph{affine} symmetry algebra (whereas global Ward identities only make use of the global subalgebra). In the unflowed sector this fact does not lead to further constraints for affine primary fields. However, as realised in \cite{Eberhardt:2019ywk}, the situation is different in the spectrally flowed sector. As we have mentioned in Section~\ref{subsec:review}, unflowed vertex operator depend on less quantum numbers because $j=h=\bar{h}$. This is compensated in the flowed sector by the existence of local Ward identities that allow for non-vanishing solutions whenever \cite{Eberhardt:2019ywk}\footnote{The same bound was derived in \cite{Maldacena:2001km} by different techniques.}
\be
\sum_{i=1}^4 w_i \geq 2 \max_{i=1, \dots 4}(w_i) -2 \ . 
\label{4ptf-bound}
\ee
When the bound \eqref{4ptf-bound} is obeyed, one can derive recursion relations in the spacetime conformal weights $(h_i,\bar{h}_i)$ for the correlators 
\be 
\left\langle \prod_{i=1}^4 V^{w_i}_{j_i, h_i}(x_i; z_i) \right\rangle \ . 
\label{VVVV}
\ee
After transforming to the $y$-basis, they become partial differential equations in the variables $x_i$, $y_i$ and $z_i$. We will not review the derivation of these constraints, it should suffice if the reader keeps in mind that there is such a local Ward identity for every spectrally flowed vertex operator in the correlator of interest. We will discuss one example in detail in Section~\ref{subsec:w=1111}.

\subsection{The Knizhnik-Zamolodchikov equation}
\label{sec:KZ-eq}

We now derive the analogue of \eqref{eq:unflowed-KZ} for correlators containing flowed vertex operators. See \cite{Giribet:2004qe,Ribault:2005ms,Giribet:2005mc,Minces:2005nb} for previous results on the KZ equation in the flowed sector. Evaluating $(k-2)L_{-1}$ on the spectrally flowed affine primary state with $w>0$ we find
\begin{align}
(k-2) \partial_{z}V_{j,h}^{w}(0;z_i) &=-2h J^3_{-1}V_{j,h}^{w}(0;z)+\sum_{n=0}^{w} J_{-1-n}^-J_n^+V_{j,h}^{w}(0;z)\\
&=-2h J^3_{-1}V_{j,h}^{w}(0;z)+\sum_{n=0}^{w-1} [J_{-1-n}^-,J_n^+]V_{j,h}^{w}(0;z)\nonumber\\
& \qquad + J^-_{-w-1}J^+_{w} V_{j,h}^{w}(0;z)+J^+_{w-1}J^-_{-w}V_{j,h}^{w}(0;z) \\
 &=-2(h-w) J^3_{-1}V_{j,h}^{w}(0;z) \nonumber  \\
 & \quad +\left(h-\tfrac{k}{2}w+j-1\right)J^+_{w-1}V_{j,h-1}^{w}(0;z)\nonumber\\
 &\quad+\left(h-\tfrac{k}{2}w-j+1 \right)J^-_{-w-1}V_{j,h+1}^{w}(0;z) \ .
 \end{align}
Translating this formula to the position $x$ leads to additional correction factors and we obtain
\begin{multline}
(k-2) \partial_{z}V_{j,h}^{w}(x;z) = -2(h-w) (J^3_{-1}-x J^+_{-1})V_{j,h}^{w}(x;z) \\ 
+\left(h-\tfrac{k}{2}w+j-1\right)J^+_{w-1}V_{j,h-1}^{w}(x;z) \\
+\left(h-\tfrac{k}{2}w-j+1\right)(J^-_{-w-1} -2x J^3_{-w-1} + x^2 J^+_{-w-1})V_{j,h+1}^{w}(x;z) \ . 
\label{eq:KZ-intermediate-1}
\end{multline}
Inserting eq.~\eqref{eq:KZ-intermediate-1} into a correlation function leads to the KZ equation. For simplicity, let us consider four-point functions and set $x=x_4$ and $z=z_4$ in \eqref{eq:KZ-intermediate-1}.\footnote{Of course, one could also set $x=x_i$ and $z=z_i$ for $i=1,2,3$. However, this is not important: all choices are equivalent by M\"{o}bius symmetry.} Applying contour deformation techniques and making use of the OPEs of the $\mathfrak{sl}(2,\mathds{R})_k$ currents with $V^w_{j,h}(x;z)$, correlators containing $J^3$ and $J^-$ modes can be rewritten in terms of correlators containing only $J^+$ modes,  
\begin{subequations}
\begin{align}
& \left\langle  \prod_{i=1}^3 V_{j_i, h_i}^{w_i}(x_i;z_i) J^3_{-m}V_{j_4,h_4}^{w_4}(x_4;z_4)  \right\rangle =  - \sum_{i=1}^3 \Biggl[ \frac{h_i + x_i \partial_{x_i}}{(z_i-z_4)^m} \left\langle \prod_{p=1}^4 V_{j_p,h_p}^{w_p}(x_p;z_p)   \right\rangle \nonumber \\
& \hspace{40pt} + \sum_{\ell=1}^{w_i} \binom{m+ \ell -2}{\ell-1} \frac{(-1)^{\ell-1} \,  x_i}{(z_i-z_4)^{m + \ell -1}} \left\langle J^+_\ell V_{j_i,h_i}^{w_i}(x_i;z_i) \prod_{p \neq i}V_{j_p,h_p}^{w_p}(x_p;z_p)   \right\rangle \Biggr] \ ,  \\
& \left\langle  \prod_{i=1}^3 V_{j_i,h_i}^{w_i}(x_i;z_i) J^-_{-m}V_{j_4,h_4}^{w_4}(x_4;z_4)  \right\rangle =  - \sum_{i=1}^3 \Biggl[ \frac{2 h_i \, x_i + x_i^2 \partial_{x_i}}{(z_i-z_4)^m} \left\langle \prod_{p=1}^4 V_{j_p,h_p}^{w_p}(x_p;z_p)   \right\rangle \nonumber\\
& \hspace{40pt} + \sum_{\ell=1}^{w_i} \binom{m+ \ell -2}{\ell-1} \frac{(-1)^{\ell-1} \, x_i^2}{(z_i-z_4)^{m + \ell -1}} \left\langle J^+_\ell V_{j_i,h_i}^{w_i}(x_i;z_i) \prod_{p \neq i}V_{j_p,h_p}^{w_p}(x_p;z_p)   \right\rangle \Biggr] \ . 
\end{align}
\label{eq:J3-Jm-correlators}
\end{subequations}
Finally, making use of the local Ward identities of \cite{Eberhardt:2019ywk}, one can solve for correlators of the form 
\be
\left\langle J^+_\ell V_{j_i,h_i}^{w_i}(x_i;z_i) \prod_{p \neq i}V_{j_p,h_p}^{w_p}(x_p;z_p)   \right\rangle \ , \qquad 0 < \ell < w_i \ ,  
\ee
in terms of correlators \eqref{VVVV}, their $h_i$ shifted values and their derivatives with respect to the $x_i$. Taking the $y$-transform we obtain a partial differential equation which is first order in $z_4, y_1, \dots, y_4$ and second order in $x_4$. Explicit examples can be found in the ancillary {\tt Mathematica} file and in Section \ref{sec:4ptf}. 

\subsection{Null vector constraints}
\label{sec:null-vectors}

An unflowed spin $j$ representation of $\mathfrak{sl}(2,\mathds{R})_k$ is reducible whenever \cite{Kac:1979fz,McElgin:2015eho}
\be 
\hspace{-20pt} j=j_{r,s}^+=\frac{1 + r +s(k-2)}{2} \ , \qquad r,s \in \mathds{Z}_{> 0}
\label{null vector spin plus}
\ee
or 
\be 
\hspace{35pt} j=j_{r,s}^-=\frac{1 - r - s(k-2)}{2} \ , \qquad r \in \mathds{Z}_{> 0} \ , \quad s \in  \mathds{Z}_{\geq 0} \ . 
\label{null vector spin minus}
\ee
The corresponding null vector $\ket{\mathcal{N}}$ appears at level $rs$ in the Verma module \cite{Malikov1984} (though as we shall explain momentarily, this does not translate into the order of the resulting differential equation). Some simple examples are
\begin{subequations}
\begin{align}
j&=j_{1,0}^-=0:&  \ket{\mathcal{N}}&=m \ket{j,m}\ , \label{eq:null-vector j=0}\\
j&=j_{2,0}^-=-\frac{1}{2}:&  \ket{\mathcal{N}}&=(2m-1)(2m+1) \ket{j,m}\ , \\
j&=j_{1,1}^-=1-\frac{k}{2}: & \ket{\mathcal{N}}&=J^-_{-1} \ket{j,m+1}-2J^3_{-1} \ket{j,m} +J^+_{-1} \ket{j,m-1}\ , \label{eq:null-vector j=1-k/2}\\
j&=j_{1,1}^+=\frac{k}{2}: & \ket{\mathcal{N}}&=(k+2 m-2) (k+2 m) J_{-1}^-\ket{j,m+1}\nonumber\\
&&&\qquad+2 (k-2 m-2) (k+2 m-2) J^3_{-1}\ket{j,m}\nonumber\\
&&&\qquad+(k-2 m-2) (k-2 m)
   J^+_{-1}\ket{j,m-1}\ .
   \label{null4m}
\end{align}
\end{subequations}
By a null vector $m \ket{0,m}$ we mean that the vector $\ket{0,m}$ is null whenever $m \ne 0$. In other words, it is the vacuum representation. It will be convenient in the following to think about these null vectors in this way.
Null vectors imply constraints in the form of differential equations on correlation functions. Their derivation is somewhat different in the unflowed and flowed sector.

\paragraph{Unflowed sector.}  Let us start with the more conventional unflowed sector. Here, we just pick the lowest weight state of the family of null vectors for the $\mathcal{D}^+$ representation. Similar formulae can be derived for $\mathcal{D}^-$ representations. In the above examples:
\begin{subequations}
\begin{align}
j&=j_{1,0}^-=0:&  \ket{\mathcal{N}}&=J_0^+ \ket{j,j}\ , \label{null1}\\
j&=j_{2,0}^-=-\frac{1}{2}:&  \ket{\mathcal{N}}&=J_0^+J_0^+ \ket{j,j}\ , \label{null2} \\
j&=j_{1,1}^-=1-\frac{k}{2}: & \ket{\mathcal{N}}&=\left(J^-_{-1}J_0^+J_0^+ +2(k-3)J^3_{-1} J_0^++(k-3)(k-2)J^+_{-1}\right) \ket{j,j}\ ,  \label{null3}\\
j&=j_{1,1}^+=\frac{k}{2}: & \ket{\mathcal{N}}&=   J^-_{-1}\ket{j,j}\ . \label{null4}
\end{align}
\end{subequations}
Following \cite{Malikov1984}, null vectors can be written in closed-form for each choice of the spins in eqs.~\eqref{null vector spin plus} and \eqref{null vector spin minus} at the price of allowing non-integer exponents for the modes. We review this construction in Appendix \ref{app:null-vectors}. Null vectors can in turn be easily translated into differential equations. The number of currents corresponds to the order of the differential equation.

The simplest example is given by $j_4=j_{1,0}^-=0$, where the null vector \eqref{null1} gives immediately
\be 
0 = \partial_{x_4}\left\langle \prod_{i=1}^3 V^0_{j_i}(x_i;z_i) \, V^0_{j_4}(x_4;z_4) \right\rangle \ ,
\label{eq:unflowed-null-eq j=0}
\ee
with $j_4=0$.

Slightly less trivial is the case $j_4=j_{1,1}^+=\frac{k}{2}$.
Standard 2D CFT techniques imply the differential equation
\begin{multline}
0 = \left\langle \prod_{i=1}^3 V^0_{j_i}(x_i;z_i) \, [J^-_{-1}V^0_{j_4}](x_4;z_4) \right\rangle \\
= \sum_{i=1}^3 \frac{[(x_i-x_4)^2 \partial_{x_i}+2(x_i-x_4)j_i]}{z_4-z_i} \left\langle \prod_{i=1}^4 V^0_{j_i}(x_i;z_i) \right\rangle
\label{eq:unflowed-null-eq-intermediate}
\end{multline}
Making use of global Ward identities and setting $x_1=0$, $x_2=1$, $x_3=\infty$ and $x_4=x$ and similarly for $z_i$, eq.~\eqref{eq:unflowed-null-eq-intermediate} simplifies to 
\begin{multline}
\partial_{x_4} \left\langle \prod_{i=1}^4 V^0_{j_i}(x_i;z_i) \right\rangle = \frac{1}{2x(x-1)(x-z)} \Bigl[-2 j_1 \left(x^2-2 x+z\right)-2 j_2 \left(x^2-z\right)\\ +2 j_3 \left(-2 x \, z+x^2+z\right)-k \left(-2 x \, z+x^2+\right)\Bigr] \left\langle \prod_{i=1}^4 V^0_{j_i}(x_i;z_i) \right\rangle \ .
\label{eq:unflowed-null-eq j=k/2}
\end{multline}
In Appendix \ref{app:null-vectors} we present the differential equations associated to the null vectors \eqref{null2} and \eqref{null3} \cite{Teschner:1997ft}. In general --- see Appendix \ref{app:null-vectors} --- the order of the null vector equation in the unflowed sector is \cite{Malikov1984}
\be 
\text{ord}(j^\pm_{r,s})=r(2s\mp 1)  \ .
\ee

\paragraph{Flowed sector.}
In addition to unflowed representations, one also considers spectrally flowed representations. Also spectrally flowed representations contain null vectors.\footnote{The spectrally flowed image of a null vector is itself null. In fact, the image of a null vector under the spectral flow automorphism gives rise to a descendant that is at the same time the spectrally flowed image of a primary state (in the spectrally flowed sense).} 
For example, for $j=j^-_{1,1}=1-\frac{k}{2}$, the vector 
\be 
(J^-_{-w-1}V^w_{j, \, h+1})(x;z) - 2 (J^3_{-1} V^w_{j, \, h})(x;z)+(J^+_{w-1}V^w_{j, \, h-1})(x;z)
\label{eq:flowed-null-vector}
\ee
is null, in analogy to the unflowed sector \eqref{eq:null-vector j=1-k/2}.
However, the logic how these null vectors are translated into differential equations is slightly different. The primary reason for this is that the Ward identities relate correlators with different values of $h$ and thus we cannot simply restrict ourselves to one state of the null representation. Instead, we proceed as in the derivation of the KZ equation and rewrite $J^3$ and $J^-$ modes in terms of $J^+$ modes only, see eqs.~\eqref{eq:J3-Jm-correlators}. After solving the local Ward identities, this reduces to a differential equation for the correlator itself. 
For example, in the case of $j_4= j^+_{1,1} =\frac{k}{2}$, the resulting differential equation is first order (in both $x_4$ and $y_i$) because it only involves one current insertion. Thus, the order of the differential equation is the same as in the unflowed sector. This is no coincidence, since we will later relate a spectrally flowed correlator to its unflowed counterpart. We will discuss the spectrally flowed null vector equations further in Section~\ref{subsec:more constraints}.

\subsection{Summary of constraints}
To keep the reader oriented, let us summarise the various constraints. All of the constraints we discussed are partial differential equations in the variables $x_1,\dots,x_4$, $y_1,\dots,y_4$ and $z_1,\dots,z_4$. There are 6 global Ward identities which allow us to put $x_1=z_1=0$, $x_2=z_2=1$ and $x_3=z_3=\infty$. One can recover the general dependence using \eqref{eq:global Ward identities solution 4pt function}. There are as many local Ward identities as there are spectrally flowed vertex operators in the correlator. Thus assuming that $w_i >0$ for all $i$, there are four local Ward identities which we will solve below and one ends up with a function depending on two variables only. This function satisfies a further constraint due to the KZ-equation. For degenerate representations, null vectors can impose additional constraints which determine the correlator fully, up to an overall constant. In the following section, we will solve all these constraints and propose a closed-form expression for the correlator.

\section{A closed-form expression for the spectrally flowed four-point function}
\label{sec:4ptf}

We will consider four-point functions of the form
\be 
\left\langle\prod_{i=1}^4  V_{j_i,h_i}^{w_i}(x_i;z_i) \right \rangle\ .
\ee
We will show that the constraints of the recursion relations can be completely solved after transforming to $y$-space and that, as anticipated in the Introduction, knowledge of the unflowed correlator leads to a complete determination of the flowed correlator. Based on different consistency conditions, we are able to give a conjecture for the prefactor.

\subsection[The flowed correlator with \texorpdfstring{$w_1=w_2=w_3=w_4=1$}{w1=w2=w3=w4=1}]{The flowed correlator with $\boldsymbol{w_1=w_2=w_3=w_4=1}$} \label{subsec:w=1111}
Before presenting our general result for the four-point functions, let us treat the simplest case where all four vertex operators are flowed. We find it convenient to perform the computation of the correlator in the $y$-space introduced in Section~\ref{subsec:review}, since in these variables all constraints can be written as partial differential equations. We consider the following correlator:
\be 
\left\langle \prod_{i=1}^4 V_{j_i}^{w_i=1}(x_i;y_i;z_i) \right \rangle\ .
\label{V1V1V1V1}
\ee 
Even though this correlator depends on four new variables $y_1$, $y_2$, $y_3$ and $y_4$ that did not enter the unflowed correlation function, we also have four more constraints coming from the local Ward identities. We again set $x_1=z_1=0$, $x_2=z_2=1$, $x_3=z_3=\infty$, $x_4=x$ and $z_4=z$. The four local Ward identities read in this case\footnote{The reader can reproduce these equations easily using the ancillary {\tt Mathematica} file.}
\begin{subequations}
\begin{align}
0&=\Big(\left(1-y_1\right) y_1 z^2 \partial_{y_1}
   +\left(1-y_2\right) z^2\partial_{y_2}+\left(1-y_3\right) z^2\partial_{y_3}
   +\left(x^2-2 \, z \, y_4\,  x+z^2 y_4\right)\partial_{y_4}  \nonumber\\
   &\quad+\left(-k x+z j_1(1-2y_1)-z j_2-z j_3+j_4(z-2x)\right) z-x (x-z)
 z \partial_x\Big) \langle \cdots \rangle \ ,  \\
 0&=\Big(\left(1-y_1\right) (1-z)^2\partial_{y_1}+\left(1-y_2\right) y_2
   (1-z)^2\partial_{y_2}+\left(1-y_3\right)
   (1-z)^2\partial_{y_3}\nonumber\\
   &\quad+\left(x^2-2\, z \, y_4 \, x+2 \, y_4 \, x-2 x+z^2 y_4-y_4+1\right) \partial_{y_4}\nonumber\\
   &\quad+\big(k(1-x)+j_1(1-z)+ j_2(z-1)(1-2y_2)+j_3(1-z)+j_4(1+z-2x) \big) 
   (z-1)\nonumber\\
   &\quad-(x-1) (x-z) (z-1)\partial_x \Big) \langle \cdots \rangle \ , \\
   0&= \Big( \left(1-y_1\right)\partial_{y_1}+\left(1-y_2\right)
   \partial_{y_2}+\left(1-y_3\right) y_3
   \partial_{y_3}+\left(1-y_4\right) 
   \partial_{y_4}\nonumber\\
   &\quad+\left(-k-j_1-j_2+j_3(1-2y_3)-j_4\right)+(z-x)
   \partial_x\Big) \langle \cdots \rangle \ ,  \\
   0&= \Big((1-z) \left( x^2(1-z)+z(1-x)^2 \, y_1\right) \partial_{y_1}+z \left(z (1- x)^2+(x^2-z)(1-z)\,  y_2\right) \partial_{y_2} \nonumber\\
   &\quad-(1-z) z \left(z(z-1)+(x^2-2zx+z)\, y_3\right) \partial_{y_3} \nonumber\\
 &\quad + (1-z) z \, y_4 \left(x^2-2 \, z \, x+z+z(z-1) \, y_4\right)\partial_{y_4} \nonumber\\
 &\quad-(1-z) z \big(k x(1-x)-(x^2-2x+z)\, j_1- ( x^2-z)\, j_2+(x^2-2zx+z)\,(j_3+j_4) \nonumber\\
   &\quad\qquad +2 \, z(1-z) \, j_4 \, y_4\big) +(1-x) x (1-z) z (z-x) \partial_x\Big) \langle \cdots \rangle   \ ,
 \end{align}
 \label{1111-local-Ward-id}
 \end{subequations}
where $\langle\cdots\rangle$ is a shorthand notation for the correlator \eqref{V1V1V1V1}. These equations look very complicated, but we should note that they are by construction first order in all variables. They are all mutually compatible and their common solution contains an undetermined function of two variables.\footnote{We started with the six variables $y_1$, $y_2$, $y_3$, $y_4$, $x$ and $z$ and solving the recursion relations fixes the dependence on four of those.} We can take one variable to be $z$, since the equations do not contain $z$-derivatives. The other variable is taken to be
 \begin{multline} 
 c=\frac{x-2z+z^2+ z(1-z)y_2+ z(1-x) y_3-x (1-z)y_2\, y_3}{x-2x\, z+z^2-(1-x)z \, y_1-x(1-z)y_2+z(1-z)y_1 \, y_2 } \\
 \times \frac{x(x-2z+z^2)+ (1-x)z^2 \, y_1+z^2(1-z)y_4-z^2(1-z)y_1 \, y_4 }{x-2 x \, z+z^2-x(1-x) y_3-z(1-z)y_4 +z(1-z)y_3 \, y_4 }\ .
 \end{multline}
By explicitly solving these equations one finds the solution
 \begin{align}
 &\left\langle \prod_{i=1}^4 V_{j_i}^{w_i=1}(x_i;y_i;z_i) \right \rangle=(1-z)^{j_2+j_4+\frac{k}{2}} z^{j_1+j_4+\frac{k}{2}} (x-z)^{j_1+j_2+j_3+j_4-k} \nonumber\\
 &\qquad \times\left(x-2 \, x \, z+z^2-(1-x)z \, y_1-x(1-z)y_2+z(1-z)y_1 \, y_2\right){}^{-j_1-j_2+j_3-j_4} \nonumber\\
   &\qquad\times\left(x-z^2-z(1-z)y_1 \, z^2-x(1-z) y_3+(1-x)z \, y_1 \, y_3\right){}^{-j_1+j_2-j_3+j_4} \nonumber\\
   &\qquad\times\left(x-2\, z+z^2+ z(1-z)y_2+ z(1-x) y_3-x (1-z)y_2 \, y_3\right){}^{j_1-j_2-j_3+j_4} \nonumber\\
   &\qquad\times\left(x-2 \, x \, z+ z^2 -x(1-x) y_3-z(1-z)y_4 +z(1-z)y_3 \,  y_4\right){}^{-2 j_4} F(c,z)\ . 
   \label{eq:sol1111}
 \end{align}
 The recursion relations of the right-movers are analogous and their solution would lead to the same right-moving structure. The undetermined function $F(c,z)$ actually depends on $c$, $\bar{c}$, $z$ and $\bar{z}$.
 The solution \eqref{eq:sol1111} is relatively simple, compared to the complexity of the differential equations \eqref{1111-local-Ward-id}. The prefactors $z$ and $1-z$ are not fixed by the recursion relations (since the recursion relations do not contain $z$-derivatives), but are inserted for later convenience.
The function $F( c,z)$ is further restricted by the KZ-equation and possible null vector decoupling equations. Since these are conceptually not more complicated for higher values of spectral flow, we postpone their discussion to Section~\ref{subsec:more constraints}.

\subsection{The general solution of the recursion relations}
\label{sec:4pt general solution}

The solution we obtained in Section~\ref{subsec:w=1111} for $w_1=w_2=w_3=w_4=1$ is suggestive. Let us compare it to the solution of the global $\mathfrak{sl}(2,\mathds{R})$ Ward identities in the unflowed sector, which is given by (compare to eq.~\eqref{eq:global Ward identities solution 4pt function})
\begin{multline}
\left\langle \prod_{i=1}^4 V_{j_i}^{0}(x_i;z_i) \right \rangle=x_{12}^{-j_1-j_2+j_3-j_4} x_{13}^{-j_1+j_2-j_3+j_4} x_{23}^{j_1-j_2-j_3+j_4} x_{34}^{-2 j_4} F\left(\frac{x_{23} \, x_{14}}{x_{12} \, x_{34}},z\right)\ ,
\end{multline}
where $x_{ij}=x_i-x_j$ and we suppressed the other variables. For $x_1=0$, $x_2=1$ and $x_3=\infty$, the prefactor trivialises. The solution to the recursion relations has the same structure, except that we have to replace the differences $x_{ij}$ by `generalised differences' $X_{ij}(x,z,y_i,y_j)$ and include a couple of simple prefactors. The general solution of the recursion relations is given by 
\begin{multline}
\left\langle \prod_{i=1}^4 V_{j_i}^{w_i}(x_i;y_i;z_i) \right \rangle=X_\emptyset^{j_1+j_2+j_3+j_4-k} X_{12}^{-j_1-j_2+j_3-j_4} X_{13}^{-j_1+j_2-j_3+j_4} \\
\times X_{23}^{j_1-j_2-j_3+j_4} X_{34}^{-2 j_4} F\left(\frac{X_{23}X_{14}}{X_{12}X_{34}},z\right)\ ,
\label{4pt-solution-even-case-small-j}
\end{multline}
where we omitted the dependence on anti-holomorphic variables and we define
\begin{multline} 
X_I(x,z,y_1,y_2,y_3,y_4)=z^{\frac{1}{2} \delta_{\{1,4\} \subset I}}\left((1-z)^{\frac{1}{2}}(-1)^{w_1w_2+w_1w_3+w_2w_4+w_3w_4}\right)^{\delta_{\{2,4\} \subset I}}\\
\times \sum_{i \in I:\ \varepsilon_i=\pm 1} P_{\boldsymbol{w}+\sum_{i \in I} \varepsilon_i e_{i}} (x;z)\prod_{i\in I} y_i^{\frac{1-\varepsilon_i}{2}}\ . \label{eq:def XI}
\end{multline}
In this definition, $I$ can be any subset of $\{1,2,3,4\}$. We made use of the polynomial $P_{\boldsymbol{w}}(x;z)$ that was defined in Section~\ref{subsec:covering maps}. The prefactor is again inserted to simplify various expressions later. It is actually not needed presently, since we have not determined the $z$-dependence anyway. 

The definition of $X_I$ plays a central role in our story, so let us elucidate it a bit further with an example. We have for example for $\boldsymbol{w}=(1,1,1,1)$
\begin{align} 
X_{12}&=P_{(2,2,1,1)}(x;z)+P_{(0,2,1,1)}(x;z)y_1+P_{(2,0,1,1)}(x;z)y_2+P_{(0,0,1,1)}(x;z) y_1 y_2 \\
   &=\frac{1}{z(1-z)}\left(x-2 \, x \, z+z^2-(1-x)z \, y_1-x(1-z)y_2+z(1-z)y_1 \, y_2\right)\ ,
\end{align}
which appears as the first of the factors in the solution \eqref{eq:sol1111}. Let us recall that the polynomials $P_{\boldsymbol{w}}(x;z)$ that appear in this expression encode the condition on $x$ and $z$ for the existence of a branched covering map from the worldsheet to the boundary of $\mathrm{AdS}_3$. In this case, only $P_{(2,2,1,1)}(x;z)$ is interesting and if it vanishes the corresponding covering map is
\be 
\gamma(\zeta)=\frac{\zeta^2}{2\zeta-1}\ .
\ee
One should also notice that by construction $X_\emptyset$ coincides with $P_{\boldsymbol{w}}(x;z)$. It should now be clear why we included the normalisation factors in \eqref{eq:def P poly}. They are needed to generate the correct solution of the local Ward identities.

Notice that at this stage the function $F(c,z)$ may depend on the choice of spectral flow parameters. In fact, both the KZ equation and the null vector equations depend on $w_1, \dots, w_4$. However, we will see in a moment that this is not the case and that $F(c,z)$ does not actually depend on the choice of $w_1, \dots, w_4$. 

The solution \eqref{4pt-solution-even-case-small-j} is only valid when $\sum_i w_i \in 2 \mathds{Z}$ and the bound \eqref{4ptf-bound} is not saturated.  Correlators saturating this bound require extra care and will be treated separately in Section \ref{sec:boundary w}. As we already saw for the three-point function, the structure of the opposite parity correlators is different and we discuss it in Section~\ref{sec:odd parity}.

If in eq.~\eqref{4pt-solution-even-case-small-j} $w_i=0$ for some $i$, then the dependence on $y_i$ drops out. This is as it should be, since in the unflowed sector $y_i$ coincides with $x_i$ and we decided to keep the latter. 

For $w_4 = 0$, the definition \eqref{eq:def XI} reduces to the one we have given for the three-point function in \cite{Dei:2021xgh}. We shall later see that if one also specialises $j_4=0$, eq.~\eqref{4pt-solution-even-case-small-j} reduces to the even parity solution in \eqref{eq:three-point function}. In fact, one can show that the remaining function $F(c,z)$ is $c$ and $z$-independent. This is in agreement with $V^{w=0}_{j=0}(x;z)$ being the identity operator on the worldsheet. 

We should note that the `generalised differences' $X_{ij}$ satisfy some analogous relations that we would expect from ordinary differences $x_{ij}= x_i - x_j$. In fact, while $X_{ij}\ne X_{i\ell}+X_{\ell j}$, we have 
\be 
1-\frac{X_{23} X_{14}}{X_{12}X_{34}}=\frac{X_{13}X_{24}}{X_{12}X_{34}}\ ,
\ee
which is the analogous relation that the $x_{ij}$'s would satisfy. Thus, $\frac{X_{23} X_{14}}{X_{12}X_{34}}$ really behaves as a cross ratio. This implies an identity on the polynomials $P_{\boldsymbol{w}}(x;z)$ that is further discussed in Appendix~\ref{app:P identities}.

Hence, somewhat surprisingly, the structure of the flowed correlator resembles very much the one of the unflowed one. While we do not know of an analytic derivation of this result, we have checked it in {\tt Mathematica} for all choices of $(w_1,w_2,w_3,w_4)$ with $\sum w_i \le 10$ (721 possibilities). We view this as ample evidence that our conjectured solution is indeed correct. 

\subsection{More constraints and the full parity even correlator} \label{subsec:more constraints}

It turns out that the analogy between the unflowed correlator and the flowed version extends beyond what was already discussed in the previous section. One can inject the solution of the recursion relations into the KZ equation, which leads to a differential equation for the function $F(c,z)$ entering \eqref{4pt-solution-even-case-small-j}.\footnote{The fact that the KZ equation is compatible with the recursion relations is another cross-check of our analysis.} We have done this in {\tt Mathematica} for all choices of $w_i$ with $\sum_i w_i \le 10$. In all those cases, it turns out that $F(c,z)$ obeys the \emph{unflowed} KZ equation \eqref{eq:unflowed-KZ} with spins $j_i$.

Similarly, in cases where $j_i$ is degenerate, one can compute the null vector equation that the correlator obeys. As explained in Section~\ref{sec:null-vectors} this again translates into a differential equation for the function $F(c,z)$. We have checked in the cases \eqref{null1} -- \eqref{null4} that the resulting differential equation coincides with the unflowed differential equation with spins $j_i$, see Appendix \ref{app:null-vectors}.

This observation can be checked with the ancillary \texttt{Mathematica} notebook. 
In the simplest case, it is easy to understand analytically. Let us assume $w_4>0$ and consider the degenerate case of $j_4=0$. In this case, the null field takes the following form (see eq.~\eqref{eq:null-vector j=0})
\be 
(h-\tfrac{kw}{2})V_{0,h}^w(x;z)\longrightarrow y_4\partial_{y_4} V_0^w(x;y;z)\ .
\ee
The first formula is the null vector in $h$-space, which translates to the given formula in $y$-space. Hence, we obtain the differential equation
\be 
\partial_{y_4} \left\langle \prod_{i=1}^4 V_{j_i}^{w_i}(x_i;y_i;z_i) \right \rangle=0 \ . 
\ee
For $j_4=0$, we see however that the prefactor of $F(c,z)$ in the solution \eqref{4pt-solution-even-case-small-j} does not depend on $y_4$ and thus the differential equation immediately reduces to
\be 
\partial_c F(c,z)=0\ ,
\ee
which coincides with the corresponding unflowed null vector equation \eqref{eq:unflowed-null-eq j=0}. 

Since the function $F(c,z)$ seems to obey the same constraints as the unflowed correlator with spins $j_i$, we are thus motivated to \emph{identify} $F(c,z)$ with the unflowed correlator, up to a normalisation constant. We can now state our general conjecture that yields the flowed correlator once the unflowed correlator is known
\begin{multline}
\left\langle V_{j_1}^{w_1}(0;y_1;0) V_{j_2}^{w_2}(1;y_2;1) V_{j_3}^{w_3}(\infty;y_3;\infty) V_{j_4}^{w_4}(x;y_4;z) \right \rangle\\
=C(w_i;j_i;k)  |X_\emptyset|^{2(j_1+j_2+j_3+j_4-k)} \,|X_{12}|^{2(-j_1-j_2+j_3-j_4)}  |X_{13}|^{2(-j_1+j_2-j_3+j_4)}  \\
\times|X_{23}|^{2(j_1-j_2-j_3+j_4)} |X_{34}|^{-4j_4}\left\langle \prod_{i=1}^4 V_{j_i}^{w_i=0}(x_i';z_i') \right \rangle\ , \label{eq:full spectrally flowed correlator}
\end{multline}
where the locations of the unflowed correlator are $x_1'=z_1'=0$, $x_2'=z_2'=1$, $x_3'=z_3'=\infty$ and
\be 
x_4'=\frac{X_{23}X_{14}}{X_{12}X_{34}} \, \qquad z_4' = z \ .
\label{x4'}
\ee
Here, we also reinstated the dependence on the right-movers. The normalisation constant will be discussed further below.

\subsection{Parity odd case} 
\label{sec:odd parity}

One can perform a similar analysis for the parity odd case. Here the solution is naturally written in terms of $X_I$ where $I$ has an odd number of elements. The solution to the local Ward identities takes the form
\begin{multline}
\left\langle \prod_{i=1}^4 V_{j_i}^{w_i}(x_i;y_i;z_i) \right \rangle=X_{123}^{\frac{k}{2}-j_1-j_2-j_3-j_4} X_{1}^{-j_1+j_2+j_3+j_4-\frac{k}{2}} \\
\times X_{2}^{j_1-j_2+j_3+j_4-\frac{k}{2}} X_3^{j_1+j_2-j_3+j_4-\frac{k}{2}} X_{4}^{-2 j_4} F\left(\frac{X_{2}X_{134}}{X_{123}X_{4}},z\right)\ .
\label{4pt-odd-parity-solution}
\end{multline}
We have checked for $\sum_i w_i \le 10$ that \eqref{4pt-odd-parity-solution} indeed provides a solution of the Ward identities. 

One can additionally impose the KZ equation and possibly null vector equations on this solution to further constrain the unknown function $F(c,z)$. The same arguments as in the previous subsection show also that the remaining unknown function $F(c,z)$ satisfies the unflowed KZ-equation with spins $(j_1,j_2,\frac{k}{2}-j_3,j_4)$. If one of the spins is degenerate, the resulting null vector equation also coincides with the unflowed null vector equation for spins $(j_1,j_2,\frac{k}{2}-j_3,j_4)$.\footnote{Note that the null vector differential equations for spin $j_{r,s}^\pm$ and $\frac{k}{2}-j_{r,s}^\pm=j_{r,s\mp 1}^\mp$ both have order $r(2s\mp 1)$, which is a good consistency check on this assertion.} Thus, we are again motivated to identify the function $F(c,z)$ with the unflowed correlator, but with spins $(j_1,j_2,\frac{k}{2}-j_3,j_4)$. We thus extend our proposal for the correlation functions to the parity odd sector as follows
\begin{multline}
\left\langle V_{j_1}^{w_1}(0;y_1;0) V_{j_2}^{w_2}(1;y_2;1) V_{j_3}^{w_3}(\infty;y_3;\infty) V_{j_4}^{w_4}(x;y_4;z) \right \rangle \\
=C(w_i;j_i;k)  |X_{123}|^{2(\frac{k}{2}-j_1-j_2-j_3-j_4)} |X_{1}|^{2(-j_1+j_2+j_3+j_4-\frac{k}{2})} \\
\times |X_{2}|^{2(j_1-j_2+j_3+j_4-\frac{k}{2})} |X_3|^{2(j_1+j_2-j_3+j_4-\frac{k}{2})} |X_{4}|^{-4 j_4} \\
\times\left\langle V_{\frac{k}{2}-j_3}^{w_3=0}(x_3';z_3')\prod_{i=1,2,4} V_{j_i}^{w_i=0}(x_i';z_i') \right \rangle\ . \label{eq:full spectrally flowed correlator parity odd}
\end{multline}
The insertion points are given by $x_1'=z_1'=0$, $x_2'=z_2'=1$, $x_3'=z_3'=\infty$ and
\be 
x_4'=\frac{X_{2}X_{134}}{X_{123}X_{4}} \ , \qquad z_4' = z \ .
\ee
Our formula looks asymmetric in the four insertion points, but as we shall discuss below, this is actually not the case.

\subsection{Four-point functions saturating the bound on spectral flow}
\label{sec:boundary w}

As already anticipated in Section \ref{sec:4pt general solution}, four-point functions saturating the bound \eqref{4ptf-bound} require particular care. It is the aim of this section to discuss them in detail. We will refer to this case in the following as the edge case.

\paragraph{Local Ward identities.} There are several interesting things happening in this case. At a technical level, it turns out that there is one more local Ward identity compared to the generic case. Hence the correlator is much more constrained than generically. Let us explain how this happens at the simple example $\boldsymbol{w}=(0,0,0,2)$. One derives the recursion relations in this case as discussed in \cite{Eberhardt:2019ywk}. One considers an insertion of a current $J^a(z)$ inside the correlator. These insertions are determined up to a finite number of terms by holomorphicity. In this case, there is only one unknown term -- the action of the mode $J_1^+$ on the fourth field. 
By imposing the various OPEs one can then derive two equations on the correlator. One is used to eliminate the unknown and the remaining one is the local Ward identity that determines the $y_4$ behaviour. In this special case, the unknown action of $J_1^+$ on the fourth field drops actually completely out when deriving the local Ward identities. Thus there is no need to eliminate it and one ends up with two linearly independent constraints on the correlator. This can be seen directly using the provided \texttt{Mathematica} code. 

Counting variables, we therefore expect that the solution of the local Ward identities only contains a function of a single cross-ratio, say $z=z_4$.  One finds that the general solution of the local Ward identities takes the form
\begin{multline} 
\left\langle V_{j_1}^{w_1}(0;y_1,0)V_{j_2}^{w_2}(1;y_2,1)V_{j_3}^{w_3}(\infty;y_3,\infty)V_{j_4}^{w_4}(x,y_4;z)\right\rangle\\
= X_{13}^{-j_1+j_2-j_3+j_4} X_{23}^{j_1-j_2-j_3+j_4} X_{12}^{2 j_3-k} X_{34}^{j_1+j_2+j_3-j_4-k} X_{1234}^{-j_1-j_2-j_3-j_4+k} f(z)\ , \label{eq:full correlator edge case}
\end{multline}
for an arbitrary function $f(z)$ (that as usual can also depend on $j_i$, $w_i$ and $k$). This form can already almost be guessed from the form of the generic even-parity four-point function \eqref{eq:full spectrally flowed correlator}. Eq.~\eqref{eq:full spectrally flowed correlator} is no longer well-defined in the edge case because $X_\emptyset=P_{\boldsymbol{w}}(x;z)=0$, since $H_{\boldsymbol{w}}<0$, see eq.~\eqref{Hurwitz-number}. Furthermore, the `generalised crossratio' degenerates as
\be 
c=\frac{X_{23}X_{14}}{X_{12}X_{34}} \longrightarrow z_4\ .
\ee
These two features can be combined to obtain still a well-defined result. Using the general relation (that makes sense away from the edge case and is discussed in Appendix~\ref{app:P identities})
\be 
z_4-c=z_4-\frac{X_{23}X_{14}}{X_{12}X_{34}}=\frac{X_\emptyset X_{1234}}{X_{12}X_{34}}\ ,
\ee
we see that we can set
\be 
F(c,z)=f(z) |c-z|^{-2j_1-2j_2-2j_3-2j_4+2k}
\ee
in \eqref{4pt-solution-even-case-small-j} to cancel the $X$-dependence and obtain a result that continues to make sense in the edge case. This then reproduces \eqref{eq:full correlator edge case}. 

This then also suggests what the function $f(z)$ should be. $F(c,z)$ was identified with the corresponding unflowed correlator and thus
\be 
f(z)=\lim_{x \to z} |x-z|^{2j_1+2j_2+2j_3+2j_4-2k}\left\langle V_{j_1}(0;0)V_{j_2}(1;1)V_{j_3}(\infty;\infty)V_{j_4}(x;z)\right\rangle\ , \label{eq:edge case f definition}
\ee
up to a possible overall constant. It is a consequence of the unflowed KZ-equation that the unflowed correlator has precisely this type of singularity so that the limit is well-defined, see \cite{Teschner:1997ft, Maldacena:2001km} and our discussion in Section~\ref{subsec:unflowed four point function}. 
\paragraph{Further constraints.} When computing the spectrally flowed KZ-equation, it turns out that the unknown action of the mode $J_1^+$ on the fourth field (in the above example $\boldsymbol{w}=(0,0,0,2)$) now enters the computation. Thus the KZ-equation does not lead to a constraint on $f(z)$. This should not come as a surprise in view of the identification, since the $x$-value of the unflowed correlator is pinned to $z$. Hence also the unflowed KZ-equation does not give any information on $f(z)$. To get a non-trivial equation on $f(z)$ we have to assume that one of the fields is degenerate in which case there is also a null vector equation. The two equations both involve the unknown mode $J_1^+$ on the fourth field that we were discussing. Hence this mode can be eliminated and we obtain a new constraint on $f(z)$. This constraint coincides with the constraint that we would find in the unflowed sector when combing the null vector equation and the KZ-equation such that the $x$-derivatives are eliminated. This gives further evidence for the identification \eqref{eq:edge case f definition}.

Recall that the KZ-equation generically predicts one out of two possible behaviours if $x$ approaches $z$. Either the correlator behaves singularly as discussed above or it behaves regularly, see eqs.~\eqref{unflowed-singular} and \eqref{delta-exp-sing-sol}. We pick out the singular behaviour in the edge case. There are however some degenerate fields for which only one of the two channels is present. Among them is in particular the identity field. Putting $j_4=0$ leads to the null vector equation $\partial_{c} F(c,z)=0$ on the unflowed correlator, which together with the KZ-equation leads to $F(x,z)=\text{const}$. This is obviously not singular as $x \to z$ and hence the edge case correlator vanishes when one of the fields is the identity. This explains why there is no non-vanishing three-point function with e.g.~$\boldsymbol{w}=(0,0,2)$ (that could be obtained from the four point function with $\boldsymbol{w}=(0,0,2,0)$ by specifying $j_4=0$). 

Contrary to the generic case of eq.~\eqref{eq:full spectrally flowed correlator}, the $y$-dependence of the edge case is fully contained in its prefactor. Hence it is in principle possible to compute the integrals over $y$-space \eqref{eq:definition y basis} to obtain the $h$-dependence of the correlators. We have not attempted to do this calculation.

\section{Further properties and consistency checks}
\label{sec:checks}

We have shown in the previous section that our conjecture \eqref{eq:main conjecture even} and \eqref{eq:main conjecture odd} follows by imposing all the symmetry constraints, up to an overall constant. As we mentioned in the Introduction, we conjecture that the missing constant is 1 in the parity even-case and $\mathcal{N}(j_3)$ in the parity odd case. In the following we will explain all further consistency checks that we performed on our proposal that give us high confidence for its correctness.

\subsection{Reduction to the three-point function}

Let us explain the reduction to our formula for three-point functions that we recalled in eq.~\eqref{eq:three-point function}. For this, we set $j_4=w_4=0$, which sets the fourth field equal to the identity field. In this case, the unflowed four-point function that enters the formula is just a three-point function and hence equal to $D(j_1,j_2,j_3)$ or $D(j_1,j_2,\frac{k}{2}-j_3)$ for the parity even and odd case, respectively. Here, $D(j_1,j_2,j_3)$ are the structure constants, see eq.~\eqref{eq:definition D}. Next, we should explain how our definition of $P_{\boldsymbol{w}}(x;z)$ reduces to the one for three-point functions. For $w_4=0$, the Hurwitz number vanishes and so by definition $\tilde{P}_{\boldsymbol{w}}(x;z)=1$. Hence the definition of $P_{\boldsymbol{w}}(x;z)$ reduces to the prefactor in eq.~\eqref{eq:def P poly}. It is also easy to see that the $x$- and $z$-dependent part of the prefactor in \eqref{eq:def P poly} are trivial and we simply have $P_{\boldsymbol{w}}(x;z)=f(\boldsymbol{w})$, where the function $f(\boldsymbol{w})$ is given in Appendix~\ref{app:function fw}. For $w_4=0$, this function further simplifies and can be written as in \cite{Dei:2021xgh}. This shows that the present definition of $X_I$ for $I \subset \{1,2,3\}$ coincides with the one we gave there. Finally, every $X_I$ for $4 \in I$ drops out of the formula because $j_4=0$ and thus one obtains \eqref{eq:three-point function} after using in the parity odd case the symmetry property
\be 
\mathcal{N}(j_1) D(\tfrac{k}{2}-j_1,j_2,j_3)=\mathcal{N}(j_2) D(j_1,\tfrac{k}{2}-j_2,j_3)=\mathcal{N}(j_3) D(j_1,j_2,\tfrac{k}{2}-j_3)\ ,
\ee
whose generalisation to the four-point function we will discuss in Section~\ref{subsec:exchange symmetry}.
\subsection{Exchange symmetry} \label{subsec:exchange symmetry}
Let us discuss another non-trivial consistency check. Obviously, the correlator in the $h$-basis has bosonic statistics and satisfies e.g.
\begin{multline}
\left\langle V_{j_1,h_1,\bar{h}_1}^{w_1}(x_1;z_1)V_{j_2,h_2,\bar{h}_2}^{w_2}(x_2;z_2)V_{j_3,h_3,\bar{h}_3}^{w_3}(x_3;z_3)V_{j_4,h_4,\bar{h}_4}^{w_4}(x_4;z_4)\right \rangle 
\\
=\left\langle V^{w_3}_{j_3,h_3,\bar{h}_3}(x_3;z_3)V^{w_2}_{j_2,h_2,\bar{h}_2}(x_2;z_2)V^{w_1}_{j_1,h_1,\bar{h}_1}(x_1;z_1)V^{w_4}_{j_4,h_4,\bar{h}_4}(x_4;z_4)\right \rangle\ ,
\end{multline}
and similarly for the other fields. In our conjecture \eqref{eq:main conjecture even} and \eqref{eq:main conjecture odd}, it is highly non-obvious that this property is satisfied, because the fields enter in an asymmetric way into the formula. This is in particular true for the parity odd case, where the third field enters in the unflowed correlator with spin $\frac{k}{2}-j_3$, whereas the others enter with spin $j_i$. 

Let us explain the exchange symmetry for the first and third field. Using global Ward identities, it corresponds to the following statement in the $y$-basis
\begin{align}
&\left\langle V_{j_1}^{w_1}(0;y_1;0)V_{j_2}^{w_2}(1;y_2;1)V_{j_3}^{w_3}(\infty;y_3;\infty)V_{j_4}^{w_4}(x;y_4;z)\right \rangle \nonumber \\
&\qquad=|x|^{-4h_4^0} |z|^{-4\Delta_4^0}\Big\langle V_{j_3}^{w_3}(0;(-1)^{w_3-1}y_3;0)V_{j_2}^{w_2}(1;(-1)^{w_2-1}y_2;1)\nonumber\\
&\qquad\qquad\qquad\qquad V_{j_1}^{w_1}(\infty;(-1)^{w_1-1}y_1;\infty)V_{j_4}^{w_4}\left(x^{-1};(-1)^{w_4-1}y_4x_4^{-2}z^{2w_4};z^{-1}\right)\Big \rangle\ ,
\end{align}
where $h_4^0$ and $\Delta_4^0$ were introduced in \eqref{eq:h0 Delta0}. We checked in \texttt{Mathematica} that in order for our conjecture to be consistent with this exchange symmetry, the unflowed correlator that enters the formula has to satisfy the following identity
\begin{align}
&\left\langle V_{j_1}^{0}(0;0)V_{j_2}^{0}(1;1)V_{j_3}^{0}(\infty;\infty)V_{j_4}^{0}(x;z)\right \rangle\nonumber\\
&\qquad=|x|^{-4j_4}|z|^{\frac{4j_4(j_4-1)}{k-2}}\left\langle V_{j_3}^{0}(0;0)V_{j_2}^{0}(1;1)V_{j_1}^{0}(\infty;\infty)V_{j_4}^{0}\left(x^{-1};z^{-1}\right)\right \rangle\ , \\[5pt]
&\mathcal{N}(j_3)\left\langle V_{j_1}^{0}(0;0)V_{j_2}^{0}(1;1)V_{\frac{k}{2}-j_3}^{0}(\infty;\infty)V_{j_4}^{0}(x;z)\right \rangle\nonumber\\
&\qquad=\mathcal{N}(j_1)|z|^{\frac{2j_4(2j_4-k)}{k-2}}\left\langle V_{j_3}^{0}(0;0)V_{j_2}^{0}(1;1)V_{\frac{k}{2}-j_1}^{0}(\infty;\infty)V_{j_4}^{0}\left(xz^{-1};z^{-1}\right)\right \rangle\ , 
\end{align}
in the parity even and odd case, respectively. The first identity is a direct consequence of the global Ward identities in the unflowed sector. The second identity is much more interesting. After combining it with the global Ward identities, it reduces to
\begin{align}
&\mathcal{N}(j_3)\left\langle V_{j_1}^{0}(0;0)V_{j_2}^{0}(1;1)V_{\frac{k}{2}-j_3}^{0}(\infty;\infty)V_{j_4}^{0}(x;z)\right \rangle\nonumber\\
&\qquad=\mathcal{N}(j_1)|x|^{-4j_4}|z|^{2j_4}\left\langle V_{j_3}^{0}(\infty;\infty)V_{j_2}^{0}(1;1)V_{\frac{k}{2}-j_1}^{0}(0;0)V_{j_4}^{0}\left(x^{-1}z;z\right)\right \rangle
\end{align}
This is identity \eqref{k/2-id}, which is derived in Appendix~\ref{app:identity}. This identity arises in the unflowed sector thanks to the existence of the degenerate representation with $j=j_{1,1}^+=\frac{k}{2}$ and doesn't follow immediately from Ward identities. We also remark that the prefactor $\mathcal{N}(j_3)$ is necessary to obtain a symmetric answer in the spins. 
Exchanges of other spins are similar to the parity even case that we discussed. They all follow from global Ward identities in the unflowed sector, which we check directly in the ancillary \texttt{Mathematica} file.

Finally, the reader might wonder why one of the spins is swapped ($j \to \frac{k}{2}-j$) in the parity odd sector and none in the parity even sector. It turns out that one can write also alternative formulae for our correlators where for the even-parity sector any even number of spins has been replaced by $j \to \frac{k}{2}-j$ and for the odd-parity sector any odd number of spins has been swapped. For each swapped spin $j_i \to \frac{k}{2}-j_i$, one has to include a prefactor $\mathcal{N}(j_i)$ in the formula. For example, the identity \eqref{k/2-id} can be rewritten as
\begin{multline}
\left\langle V^0_{j_1}(0;0) V^0_{j_2}(1;1) V^0_{j_3}(\infty;\infty) V^0_{j_4}(x;z) \right\rangle \\
= \mathcal N(j_1)  \mathcal N(j_3) |x|^{-4j_4} |z|^{2j_4} \left\langle V^0_{\frac{k}{2}-j_1}(0;0) V^0_{j_2}(1;1) V^0_{\frac{k}{2}-j_3}(\infty;\infty) V^0_{j_4}\Bigl(\frac{z}{x};z \Bigr) \right\rangle \ .  
\end{multline}
Since the left-hand side enters our formula \eqref{eq:main conjecture even} in the even-parity case, we could alternatively also use the right hand side to express it in terms of the correlator where both $j_1$ and $j_3$ have been swapped. Similarly we could swap two other spins or all four spins. Similar reasoning applies in the parity odd sector. 

\subsection{Four-point functions and reflection symmetry} \label{subsec:four point function reflection symmetry}
We will now show that the reflection symmetry of the spectrally flowed correlator for continuous representations is a consequence of properties of the unflowed correlator that enters \eqref{eq:main conjecture even} and \eqref{eq:main conjecture odd}.

\paragraph{Parity even case.} Let us discuss the reflection symmetry first in the parity even case (i.e.~$\sum_i w_i \in 2\mathds{Z}$). We also restrict to the typical case where $\sum_i w_i \ge 2\max_i w_i$. The edge case follows once we interpret it correctly as a limiting case of the typical case. In this case, the four-point function is given by eq.~\eqref{eq:full spectrally flowed correlator}. To demonstrate reflection symmetry, we work directly in the $y$-basis, where reflection symmetry acts according to \eqref{eq:y-basis-reflection}. Let us reflect $j_4$. This calculation implies that a similar result is true also for $j_1$, $j_2$ and $j_3$ because by the exchange symmetry that we discuss in Section~\ref{subsec:exchange symmetry}, we can let any field play the role of the fourth field.

So we want to compute the integral
\be 
\int \mathrm{d}^2 y_4 \ |y-y_4|^{4j_4-4} |X_{34}|^{-4j_4} F \left(\frac{X_{23}X_{14}}{X_{12}X_{34}},z\right)\ ,
\ee
where we omitted all factors in \eqref{eq:full spectrally flowed correlator} that do not depend on $y_4$ and $F(x,z)$ denotes the unflowed four-point function. One can explicitly compute this integral by changing variables to 
\be 
c=\frac{X_{23}X_{14}}{X_{12}X_{34}}\ .
\ee
One gets
\be 
|AD-BC|^{2-4j_4} |X_{34}(y)|^{4j_4-4} \int \mathrm{d}^2 c \ \left|c-\frac{A +B y}{C+D y} \right|^{4j_4-4} F(c,z)\ ,
\ee
where $X_{34}(y)$ is $X_{34}$ with $y_4$ replaced by $y$ and
\be 
A+B y_4=\frac{X_{23}X_{14}}{X_{12}}\ , \quad C+D y_4=X_{34}\ ,
\ee
so that $A$, $B$, $C$ and $D$ are $y_4$-independent. The quadratic identity \eqref{eq:2nd quadratic identity} yields
\begin{align}
AD-BC=\pm \frac{X X_{13}X_{23}}{X_{12}}\ .
\end{align}
Moreover, 
\be 
\frac{A +B y}{C+D y}=\frac{X_{23}X_{14}(y)}{X_{12}X_{34}(y)}
\ee
is the `new' cross ratio that depends now on $y$ instead of $y_4$. At this point the integral over $c$ is exactly the integral that one would compute for the unflowed correlator to obtain the reflected correlator \eqref{eq:y-basis-reflection}, see eq.~\eqref{eq:Teschner-reflection}. Since the unflowed correlator is by assumption reflection symmetric, the integral just evaluates to the reflected correlator (with reflection coefficient). The extra factors of $X$ and $X_{ij}$ that we obtained through the change of variables is exactly what is needed to also turn around the $j_4$ in the prefactors of eq.~\eqref{eq:full spectrally flowed correlator}. Thus we conclude that reflection symmetry of the flowed correlators follows directly from the reflection symmetry of the unflowed correlators.

\paragraph{Parity odd case.} The parity odd case is very similar. Essentially the same calculation reduces the check of reflection symmetry for the flowed correlator to the reflection symmetry of the unflowed correlator. In this case, one changes variables to
\be 
c=\frac{X_2X_{134}}{X_{123}X_4}\ .
\ee
and uses the quadratic identity \eqref{eq:3rd quadratic identity}.

\paragraph{Edge case.} We already mentioned that the edge case (i.e.\ when the bound \eqref{4ptf-bound} is saturated) can be understood as a limiting case of the parity even case. Let us be more precise here. In the edge case, we found that the solution was expressed in terms of the limit of the unflowed correlator \eqref{eq:edge case f definition}
\be 
f_{j_4}(z)=\lim_{x \to z}|x-z|^{2j_1+2j_2+2j_3+2j_4-2k}\left\langle V_{j_1}(0;0)V_{j_2}(1;1)V_{j_3}(\infty;\infty)V_{j_4}(x;z)\right\rangle\ , 
\ee
where we momentarily emphasised the dependence on $j_4$ (even though $f(z)$ also depends of course on the other spins). Using \texttt{Mathematica} one can check that reflection symmetry in the flowed sector reduces to the following identity for this limit of the unflowed correlator:
\begin{align}
f_{1-j_4}(z)=R_{1-j_4}(1-2j_4) f_{j_4}(z)\frac{\lgamma(2j_4-1)\lgamma(k-j_1-j_2-j_3-j_4+1)}{\lgamma(k-j_1-j_2-j_3+j_4)} \ , \label{eq:reflection symmetry edge case}
\end{align}
where $R_j$ is the reflection coefficient \eqref{eq:reflection coefficient} and $\lgamma(x)=\Gamma(x)/\Gamma(1-x)$.
This identity follows almost directly from reflection symmetry of the unflowed correlator. It can be demonstrated as follows.
\begin{align}
f_{1-j_4}(z)&=\frac{(1-2j_4)R_{1-j_4}}{\pi} \lim_{x \to z}|x-z|^{2j_1+2j_2+2j_3+2(1-j_4)-2k}\int \mathrm{d}^2 x' \ |x-x'|^{4j_4-4} \nonumber\\
&\qquad\qquad\qquad\qquad\qquad\qquad\times\left\langle V_{j_1}(0;0)V_{j_2}(1;1)V_{j_3}(\infty;\infty)V_{j_4}(x',z)\right\rangle \\
&=\frac{(1-2j_4)R_{1-j_4}}{\pi} \int \mathrm{d}^2 u \ |u|^{4j_4-4} \lim_{x \to z}|x-z|^{2j_1+2j_2+2j_3+2j_4-2k}\nonumber\\
&\times\left\langle V_{j_1}(0;0)V_{j_2}(1;1)V_{j_3}(\infty;\infty)V_{j_4}(x+u(x-z),z)\right\rangle
\end{align}
where we changed variables $x'=x+u(x-z)$ in the integral and we assumed that it is allowed to interchange the integral with the limit. The limit yields now by definition  $|u+1|^{-2j_1-2j_2-2j_3-2j_4+2k}f_{j_4}(z)$. The remaining integral can be evaluated and leads to the combination of $\lgamma$-functions that appear in \eqref{eq:reflection symmetry edge case}. Thus, we have also checked reflection symmetry in the edge case.

\subsection{Spectrally flowed correlators \`a la Fateev-Zamolodchikov-Zamolodchikov}

Correlators with insertions of spectrally flowed vertex operators were first computed in \cite{Fateev}, see also \cite{Maldacena:2001km}. In this section we will compute the correlator 
\be 
\left\langle V^0_{j_1}(0;0) V^0_{j_2}(1;1) V^0_{j_3}(\infty;\infty) V^1_{j_4,h_4}(x;z) \right\rangle
\label{0001}
\ee
$\grave{\text{a}}$ la Fateev-Zamolodchikov-Zamolodchikov and show that the result agrees with what follows from our proposal \eqref{eq:main conjecture odd}. According to \cite{Fateev, Maldacena:2001km}, the vertex operator with $w=1$ can be defined as\footnote{The normalisation is usually neglected in the literature. It can be reinstated by taking into account the various $k$-dependent contributions in \cite{Maldacena:2001km}.}
\begin{multline}
V^{1}_{j, h, \bar h}(x;z) \equiv \sqrt{\frac{B\left(0 \right)}{B\left( \frac{k}{2} \right)}} \\
\times \, \lim_{\epsilon \to 0} \epsilon^{h-\frac{k}{2}} \epsilon^{\bar h-\frac{k}{2}} \int \text{d}^2 y \, y^{j-h+\frac{k}{2}-1} y^{j-\bar h+\frac{k}{2}-1} \,  V^0_j(x+y,z+\epsilon) V^0_{\frac{k}{2}}(x;z) \ .
\label{Fateev-spectral-flow} 
\end{multline}
Eq.~\eqref{Fateev-spectral-flow} is understood to hold inside any correlation function. In order to compute the correlator \eqref{0001} one should then consider the five-point function 
\begin{align}
& \left\langle V^0_{j_1}(0;0) V^0_{j_2}(1;1) V^0_{j_3}(\infty;\infty) V^0_{j_4}(x_4;z_4) V^0_{\frac{k}{2}}(x_5;z_5) \right\rangle \nonumber \\
& \qquad = \frac{B \left(\frac{k}{2}\right)}{B\left(\frac{k}{2}-j_3\right)}  |1-x_5|^{2(-\frac{k}{2}+j_1-j_2+j_3+j_4)} \, |x_5|^{2(-\frac{k}{2}-j_1+j_2+j_3+j_4)} \, |x_4-x_5|^{-4j_4} \nonumber \\
& \qquad \qquad \times \, |x_5-z_5|^{2(\frac{k}{2}-j_1-j_2-j_3-j_4)}  |1-z_5|^{2j_2} |z_5|^{2j_1} |z_4-z_5|^{2j_4} \nonumber \\
& \qquad \qquad \times \, \left\langle V^0_{j_1}(0;0) V^0_{j_2}(1;1) V^0_{\frac{k}{2}-j_3}(\infty;\infty) V^0_{j_4}\left(\frac{(x_5-1)(x_5 z_4 -x_4 z_5)}{(x_5-x_4)(x_5-z_5)};z_4\right) \right\rangle
\label{5ptf-unflowed-k/2}
\end{align}
and set  
\be 
x_4 = x + y \ , \qquad z_4 = z+\epsilon \ , \qquad x_5 = x \ , \qquad z_5 = z \ . 
\ee
Eq.~\eqref{5ptf-unflowed-k/2} simply follows from injecting \eqref{fusion-inf} into \eqref{KZ-null-sol}. After performing the change of variable $y \to y \epsilon$ in \eqref{Fateev-spectral-flow} and taking the $\epsilon \to 0$ limit, we obtain 
\begin{align}
& \left\langle V^0_{j_1}(0;0) V^0_{j_2}(1;1) V^0_{j_3}(\infty;\infty) V^1_{j_4,h_4}(x;z) \right\rangle \nonumber  \\
& \quad =\mathcal{N}(j_3) \, |1 - x|^{2(-\frac{k}{2} + j_1- j_2 + j_3 + j_4)} \, |x|^{2(-\frac{k}{2} - j_1 + j_2 + j_3 + j_4)}  \nonumber \\
& \ \quad \times \, |x - z|^{2(\frac{k}{2} - j_1 - j_2 - j_3 - j_4)} \, |1 - z|^{2 j_2} \, |z|^{2 j_1} \nonumber \\
& \ \quad \times \int \text{d}^2 y \, y^{\frac{k}{2}-h_4-j_4-1} \bar y^{\frac{k}{2}-\bar h_4-j_4-1} \left\langle V^0_{j_1}(0;0) V^0_{j_2}(1;1) V^0_{\frac{k}{2}-j_3}(\infty;\infty) V^0_{j_4}\Bigl(\tfrac{(1-x)(x-yz)}{y(x-z)};z_4\Bigr) \right\rangle  \ . 
\label{0001-Fateev}
\end{align}
Since 
\begin{align}
X_1 &= \frac{x}{\sqrt{z}} \ , & X_2 &= \frac{1-x}{\sqrt{1-z}} \ , &  X_3 &= 1 \ , \\
X_4 &= y_4 \ , & X_{123} &= \frac{z-x}{\sqrt{z(1-z)}} \ , & X_{134} &= \frac{yz-x}{\sqrt{z}} \ ,  
\end{align}
eq.~\eqref{0001-Fateev} exactly reproduces what one would expect from our proposal \eqref{eq:main conjecture odd}. 

\subsection{Coincidence limit and spectral flow violation}

In this section, we make contact with previous results in the literature through another limiting case. In the literature mostly the limit where all $x_i$'s are either 0 or $\infty$ was considered. Here we analyse this limit on our conjectured four-point functions and show that they reduce to known correlators in the literature. As a first step, we use the solution to the global Ward identity \eqref{eq:global Ward identities solution 4pt function} to put $x_2$ at a generic position. We then consider the limit
\be 
\lim_{\genfrac{}{}{0pt}{}{x_2 \to 0}{x_4 \to 0}} \left\langle V_{j_1}^{w_1}(0;y_1;0)V_{j_2}^{w_2}(x_2;y_2;1)V_{j_3}^{w_3}(\infty;y_3;\infty)V_{j_4}^{w_4}(x_4;y_4;z_4) \right \rangle\ .
\ee
This limit is generically singular. However there are some exceptions where one can take the limit. They occur for
\be 
|w_1+w_2+w_4-w_3| \le 2\ .
\ee
In the cases $|w_1+w_2+w_4-w_3| \le 1$ this is quite straightforward, whereas the two edge cases are more subtle. 
\paragraph{Spectral flow conserving correlator.} Let us start by discussing the spectral flow conserving correlator, i.e.\ the correlator satisfying $w_3=w_1+w_2+w_4$. After a rescaling of the $y_i$ and the change of variables $y_3 \to - y_3^{-1}$, we find that the correlator takes the following simple form in the $h$-basis (omitting the right-moving $y$-dependence):
\begin{align}
&\left\langle V_{j_1,h_1}(0;0)V_{j_2,h_2}(0;1)V_{j_3,h_3}(\infty;\infty)V_{j_3,h_3}(0;z) \right \rangle=(-1)^{w_1 h_1+w_3 h_2+h_3+(w_1+w_4) h_4}\nonumber\\
&\qquad\times z^{\frac{k w_1 w_4}{2}-w_1h_4-w_4h_1}(1-z)^{\frac{k w_2 w_4}{2}-w_2h_4-w_4h_2}\nonumber\\
&\qquad\times\int \prod_{i=1}^4 \mathrm{d} y_i \prod_{i=1,2,4} y_i^{\frac{k w_i}{2}-h_i+j_i-1}y_3^{\frac{k w_3}{2}+h_3+j_3-1}   y_{12}^{-j_1-j_2+j_3-j_4}y_{13}^{-j_1+j_2-j_3+j_4} y_{23}^{j_1-j_2-j_3+j_4}\nonumber\\
&\qquad\qquad\times y_{34}^{-2j_4} \left\langle V_{j_1}^0(0;0) V_{j_2}^0(1;1) V_{j_3}^0(\infty;\infty) V_{j_4}^0\left(\frac{y_{32} \, y_{41}}{y_{21}\, y_{34}};z\right) \right \rangle \ . \label{eq:coincidence limit spectral flow preserving}
\end{align}
From our perspective, it is somewhat surprising that this expression depends in a very simple way on the spectral flow --- the dependence is fully contained in the prefactor. We also notice that the $y$-dependence of the integral just comes from the usual global Ward identities and we could write the integrand as $\left\langle V_{j_1}^0(y_1;0) V_{j_2}^0(y_2;1) V_{j_3}^0(y_3;\infty) V_{j_4}^0(y_4;z) \right \rangle$. Hence it might be more appropriate to rename $y_i \to x_i$ in this context. The integral then simply is the transform from the $x$-basis to a basis where $J_0^3$ is diagonalised. This is exactly the same change of basis as going from the $y$-basis to the $h$-basis. Finally we notice that we can write the prefactor as a ratio of two free boson correlators with momenta 
\be 
m_i=\left(h_i-\frac{k w_i}{2}\right) \ . 
\ee
Then the prefactor takes the form
\be 
\frac{z^{\frac{2m_1 m_4}{k}}(1-z)^{\frac{2m_2m_4}{k}}}{z^{\frac{2h_1 h_4}{k}}(1-z)^{\frac{2h_2h_4}{k}}}\ .
\ee
This form can be derived by decomposing the algebra according to $\mathfrak{sl}(2,\mathds{R})_k \supset \frac{\mathfrak{sl}(2,\mathds{R})_k}{\mathfrak{u}(1)} \times \mathfrak{u}(1)$ and using the fact that spectral flow only acts on the $\mathfrak{u}(1)$ part \cite{Ribault:2005ms}. 

There is one further qualitative difference of these limiting correlators that doesn't occur in general. We notice that the integrand is homogeneous under rescaling $y_i \to \lambda y_i$ with homogeneity degree
\be 
h_1+h_2+h_4-h_3=m_1+m_2+m_4-m_3\ .
\ee
Thus, the integral has to vanish except if this condition is true, which leads to momentum conservation in the boson prefactors. This recovers \cite[eq.~(2.23)]{Ribault:2005ms} and serves as a consistency check of our analysis.

\paragraph{Spectral flow violation by two units.} Next we consider the maximally spectral flow violating case. Let us consider the case $w_1+w_2+w_4=w_3-2$, which corresponds to the edge case of the correlators discussed in Section~\ref{sec:boundary w}. The case with $w_1+w_2+w_4=w_3+2$ is again identical after the replacement $w_3\to -w_3$ has been performed. As explained there only a limit of the unflowed correlator enters in these correlators. As a consequence, the integral over $y_i$ becomes completely explicit and beyond the prefactors already present in \eqref{eq:coincidence limit spectral flow preserving} we obtain (omitting right-moving dependence)
\begin{align}
&(-1)^{h_1+h_2+h_3}z^{h_2-j_2-\frac{k w_2}{2}-h_3-j_3+\frac{k w_3}{2}}(1-z)^{h_1-j_1-\frac{k w_1}{2}-h_3-j_3+\frac{k w_3}{2}} \nonumber\\
&\qquad\times\int \mathrm{d} y_i \ \prod_{i=1,2,4} y_i^{\frac{k w_i}{2}-h_i+j_i-1}y_3^{-\frac{k w_3}{2}+h_3+j_i-1} (y_1+y_2+y_3+y_4)^{k-j_1-j_2-j_3-j_4} \nonumber\\
&\qquad\times\lim_{x \to z} (x-z)^{j_1+j_2+j_3+j_4-k} \langle V_{j_1}^0(0;0) V_{j_2}^0(1;1) V_{j_3}^0(\infty;\infty)V_{j_4}(x;z) \rangle\ .
\label{violationw2}
\end{align}
Of course for this limit to make sense we should rather use the non-chiral analogue of this formula. One can in principle evaluate this formula further because the $y$-integral is simple to compute, but we did not find it very insightful to do so. 
A closed-form formula for correlators with maximal spectral flow violation (2 in our case) in terms of Liouville correlators has been proposed in \cite{Ribault:2005ms} and confirmed in \cite{Giribet:2011xf}. 
One can show the equivalence with the formula presented here as follows. We have \cite{Ponsot:2002cp}
\begin{multline}
\lim_{x \to z} |x-z|^{2(j_1+j_2+j_3+j_4-k)} \langle V_{j_1}^0(0;0) V_{j_2}^0(1;1) V_{j_3}^0(\infty;\infty)V_{j_4}(x;z) \rangle\\
=C \, \lgamma(j_1+j_2+j_3+j_4-k) |z|^{2j_2+2j_3-k}|1-z|^{2j_1+2j_3-k}\langle V_{\alpha_1}(0)V_{\alpha_2}(0)V_{\alpha_3}(0)V_{\alpha_4}(z) \rangle_\text{L}\ ,
\end{multline}
where
\begin{align}
C&=-\frac{\nu^{-2-b^{-2}}}{2\pi^2 b^7 \lgamma(b^2)^2}\ ,\\
\alpha_i&=b+\frac{1}{2b}-b j_i
\end{align}
and $b^{-2}=k-2$. The correlator on the right hand side is a correlator in Liouville theory. This can be derived as a special case from the $H_3^+$-Liouville correspondence \cite{Ribault:2005wp}. We follow the standard conventions for Liouville theory \cite{Zamolodchikov:1995aa}. The Liouville cosmological constant $\mu_\text{L}$ is related to the parameter $\nu$ as follows,
\be 
\mu_\text{L}=\frac{1}{\pi \nu\,  \lgamma(b^2)}\ .
\ee
Evaluating the integral above over the $y$-coordinates and assuming continuous representations leads to 
\begin{align}
&\lim_{\genfrac{}{}{0pt}{}{x_2 \to 0}{x_4 \to 0}} \left\langle V_{j_1}^{w_1}(0;y_1;0)V_{j_2}^{w_2}(x_2;y_2;1)V_{j_3}^{w_3}(\infty;y_3;\infty)V_{j_4}^{w_4}(x_4;y_4;z_4) \right \rangle \nonumber\\
&\qquad=\pm i (2\pi)^2 C\delta^{(2)}(m_3-m_1-m_2-m_4+k)  \, z^{\beta_{14}}\bar{z}^{\bar{\beta}_{14}} \, (1-z)^{\beta_{24}}(1-\bar{z})^{\bar{\beta}_{24}} \nonumber \\
&\qquad\qquad\times\prod_{i=1,2,4} \lgamma(j_i-m_i) \, \lgamma(m_3-j_3)\ \langle V_{\alpha_1}(0)V_{\alpha_2}(0)V_{\alpha_3}(0)V_{\alpha_4}(z) \rangle_\text{L}\ ,
\end{align}
where $m_i=h_i-\frac{kw_i}{2}$ and
\be 
\beta_{ij}=\frac{k}{2}(1-w_iw_j)-m_i(w_j+1)-m_j(w_i+1)\ .
\ee
Here, we suppressed the signs that arose as prefactors since they are convention dependent anyway. This formula matches precisely with \cite{Ribault:2005ms} and \cite{Giribet:2011xf}.\footnote{To match their convention one has to replace $w_3$ by  $-w_3$ and $j_3$ by $-j_3$. This is because we inserted the third field at $x_3=\infty$ and used positive spectral flow, whereas in \cite{Ribault:2005ms} and \cite{Giribet:2011xf} all fields are inserted at $x_i=0$ and instead a negative amount of spectral flow is used. The two pictures are completely equivalent.}
 Our result also gives a value for the unknown prefactor in their formula.

\paragraph{Spectral flow violation by one unit.} Finally, we look at the correlators that violate spectral flow by one unit. In this case, we find the same almost trivial dependence of the correlator on spectral flow as in the spectral flow conserving case. We get the same prefactor in front of the integral, whereas the integral itself takes the form (after a change of variables)
\begin{align}
&\int \prod_{i=1}^4 \mathrm{d}^2 y_i \ \prod_{i=1,2,4} y_i^{-\frac{k w_1}{2}+h_i+j_i-1}  y_3^{\frac{k w_3}{2}-h_3+j_3-1} (y_1+y_2+y_3)^{\frac{k}{2}-j_1-j_2-j_3-j_4}  \nonumber\\
&\qquad\times \left\langle V_{j_1}^0(0;0) V_{j_2}^0(1;1) V_{j_3}^0(\infty;\infty) V_{j_4}^0\left(\frac{y_1+z y_3- y_4}{y_1+y_2+y_3};z\right) \right \rangle\ .
\label{violationw1}
\end{align}
for $w_1+w_2+w_4=w_3+1$. The integrand for $w_1+w_2+w_4=w_3-1$ is identical, except that the replacement $w_i \to -w_i$ has to be performed. We again note that the integrand is homogeneous in a joint rescaling of the $y_i$'s. This implies again the conservation of the $J_0^3$-eigenvalues
\be 
h_1+h_2+h_4=h_3\ .
\ee
The correlator violating spectral flow conservation by one unit has been computed in \cite{Ribault:2005ms} in terms of correlators of Liouville theory. Contrary to the maximally spectral flow violating case, the Liouville correlator involves a five-point function and the relation of our formula to the formula in terms of Liouville correlators becomes less straightforward. 
We expect that the two results will agree, even though we have not tried to show this explicitly.  

\section{Conclusions, discussion and open questions}
\label{sec:conclusions}

In this paper we have studied correlators of the analytically continued $\text{SL}(2,\mathds{R})$ WZW model and proposed a closed-form formula for four-point functions with an arbitrary amount of spectral flow, see eq.~\eqref{eq:main conjecture}. We have shown that our proposal is consistent with all the symmetries of the model, namely global and local Ward identities, Knizhnik-Zamolodchikov equation, null vector equations (when applicable), reflection symmetry and exchange symmetry. 

\medskip 

Solving a 2D CFT requires, beyond knowledge of spectrum and structure constants, a good control over the conformal block expansion of correlators and a proof of crossing symmetry. Usually, formulae for four-point functions in 2D CFTs are expressed in terms of conformal block expansions. However, in eq.~\eqref{eq:main conjecture} our logic is different --- we would first determine the unflowed correlator through a conformal block expansion that is well-understood thanks to the work of Teschner \cite{Teschner:1997ft}. After the unflowed correlator has been found the flowed one follows through the integral transform that we have given. Nonetheless, it would be interesting to understand whether one can obtain a conformal block expansion directly in the flowed sector. In fact, we feel that the OPE of spectrally flowed vertex operators deserves a better understanding. A related issue is crossing symmetry of the $\text{SL}(2,\mathds{R})$ WZW model. This consistency requirement has been shown only in the unflowed sector \cite{Teschner:2001gi} for the $H_3^+$ model. Since eq.~\eqref{eq:main conjecture} gives the flowed four-point function in terms of the unflowed correlator, one might be able to prove crossing symmetry in the flowed sector by deducing it from the unflowed sector.  

We saw that consistency of our construction implies a number of highly non-obvious identities for the irreducible polynomials $P_{\boldsymbol{w}}(x;z)$ that encode the geometry of branched covers that we have collected in Appendix~\ref{app:P identities}. As far as we are aware these identities are unknown in the mathematical literature and for a further understanding a direct proof of them would be beneficial. We also suspect that similar identities hold for the polynomials that encode higher-point correlation functions.\footnote{In general, there are $n-3$ polynomials that encode the relation of the $n-3$ crossratios in $x$ and $z$ space.} However, we guessed the prefactors of the polynomials directly which does not give us much intuition how this should generalise. 

Knowing the higher-point generalisation of the polynomials $P_{\boldsymbol{w}}(x;z)$ that we mentioned in the previous paragraph would be a prerequisite to a generalisation of our formula to arbitrary $n$-point functions. We suspect that a similar generalisation even exists for higher genus correlators, where the polynomials $P_{\boldsymbol{w}}(x;z)$ should be appropriately generalised to higher genera. Thus, we think that our formula scratches only the tip of the iceberg and the relation between unflowed and flowed correlators hides an intriguing  mathematical structure.

\medskip

As we already mentioned at various places, our study of $\text{AdS}_3$ correlators is motivated by holography. Despite various efforts \cite{Seiberg:1999xz, Argurio:2000tb, Eberhardt:2019qcl,Dei:2019osr}, the exact incarnation of the $\text{CFT}_2$ dual to strings propagating on AdS$_3$ with pure NS-NS flux is still unclear. We believe that the findings of this paper, together with \cite{Dei:2021xgh}, furnish new important indications on the nature of the dual CFT candidate. In particular, the correlators \eqref{eq:main conjecture} feature an intriguing singularity structure that we plan to explore in a future publication \cite{paper3}. Beyond the usual singularities, due to the collision of vertex operators in spacetime or on the worldsheet, we find a rich collection of singularities located at the middle of the string moduli space. These are a manifestation of the existence of worldsheet instantons as discussed in \cite{Maldacena:2001km} for the unflowed sector. The behaviour of worldsheet correlators near these singularities gives direct access to the string correlator and hence provides an important peephole to learn about the nature of the dual $\text{CFT}_2$. 

We should mention that in the regime where the string tension is minimal ($k=1$) the status of the $\text{AdS}_3/\text{CFT}_2$ duality is much better understood than what we mentioned in the previous paragraph. In recent years a series of publications collected strong evidence that tensionless string theory on $\text{AdS}_3 \times \text{S}^3 \times \mathbb{T}^4$ is exactly dual to the symmetric product orbifold of $\mathbb{T}^4$ \cite{Giribet:2018ada, Gaberdiel:2018rqv,Eberhardt:2018ouy,Eberhardt:2019ywk,Dei:2020zui,Eberhardt:2020bgq,Eberhardt:2020akk,Knighton:2020kuh,Hikida:2020kil,Gaberdiel:2020ycd,Gaberdiel:2021njm}. Beyond a complete matching of the unprotected perturbative spectrum, also the structure of the correlators has been matched. It is then natural to ask what our formula \eqref{eq:main conjecture} implies for the tensionless string. We plan to discuss this further in a future publication \cite{paper3}, but we can already anticipate that by performing the string integral and building on previous results \cite{Dei:2020zui} we exactly reproduce symmetric orbifold correlators from the worldsheet. 

\acknowledgments 
We would like to thank Andrea Cappelli, Matthias Gaberdiel, Gaston Giribet, Sergio Iguri, Bob Knighton, Nicolas Kovensky, Juan Maldacena, Sylvain Ribault, Cumrun Vafa, Edward Witten and Xi Yin for useful discussions and correspondence. We are grateful to Sylvain Ribault for his comments on a preliminary version of this paper. 
The work of A.D. is funded by the Swiss National Science Foundation via the Early Postdoc.Mobility fellowship.  LE is supported by the IBM Einstein Fellowship at the Institute for Advanced Study. 	
\appendix

\section{An algorithm for the polynomial \texorpdfstring{$\boldsymbol{\tilde{P}_{\boldsymbol{w}}(x;z)}$}{Pw(x,z)}} \label{app:algorithm}
The polynomials $\tilde{P}_{\boldsymbol{w}}(x;z)$ feature prominently in our solution of the four-point function. In this appendix, we describe the algorithm we used to compute them. We assume here that $w_i>0$ for every $i$ and $H_{\boldsymbol{w}}>0$. If this is not the case, we fix $\tilde{P}_{\boldsymbol{w}}(x;z)$ as described in Section~\ref{subsec:covering maps}. We follow closely \cite{Pakman:2009zz}. The covering map can be written as
\be 
\gamma(\zeta)=\frac{f_1(\zeta)}{f_2(\zeta)}
\ee
for two polynomials $f_1$ and $f_2$ of degrees\footnote{Recall that we put $x_3=z_3=\infty$, so the degree of the denominator polynomial is lower.}
\be 
d_1=\frac{w_1+w_2+w_3+w_4}{2}-1\ , \qquad d_2=\frac{w_1+w_2-w_3+w_4}{2}-1\ .
\ee
Both $f_1$ and $f_2$ satisfy Heun's differential equation
\be 
f''-\left(\frac{w_1-1}{\zeta}+\frac{w_2-1}{\zeta-1}+\frac{w_4-1}{\zeta-z}\right)f'+\frac{d_1d_2\zeta+q}{\zeta(\zeta-1)(\zeta-z)}f=0\ .
\ee
Here, $q$ is the so-called accessory parameter of Heun's equation that is unfixed at this point. To find the covering map, we need to determine polynomial solutions of this equation as described in \cite{Pakman:2009zz}. This gives a recursion relation for the coefficients of the polynomial. Since $0$ is a ramification point of order $w_1$ of the map $\gamma$, we can take $f_1(\zeta)=\zeta^{w_1}+\mathcal{O}(\zeta^{w_1+1})$.
Requiring that $f_1$ has indeed degree $d_1$ in the solution determines $q$
in terms of $z$. The second independent solution is determined by requiring that it has a non-vanishing constant term and is automatically a polynomial of order $d_2$. Thus, this algorithm determines the covering map as a function of $(z,q)$ and leads also to a polynomial relation $g(z,q)=0$ between $z$ and $q$.\footnote{This sometimes yields polynomials $g$ that are not irreducible. One finds however that the true relation between $z$ and $q$ is always irreducible and it can be found by checking the factors of $g(z,q)$ individually. In our \texttt{Mathematica} notebook, the function \texttt{zqRelPre[w\_,z\_,q\_]} computes the relation between $z$ and $q$ using this algorithm and \texttt{zqRel[w\_,z\_,q\_]} picks the correct irreducible factor. We will in the following assume that $g(z,q)$ is the irreducible factor.	}

To find $\tilde{P}_w(x;z)$, one only has to combine this information. $x$ is determined to be $\gamma(z)$, which is a rational expression in $z$ and $q$. We are thus given two polynomial conditions
\be 
g(z,q)=0\ , \qquad h(x,z,q)=0\ ,
\ee
where the second condition comes from setting $x=\gamma(z)$ (and is linear in $x$). All that remains to be done is to eliminate $q$ from these two polynomial equations. Since both are non-linear in $q$, it is impracticable to solve one equation for $q$ and insert it into the other. Instead we use some basic commutative algebra to achieve this.

In the language of commutative algebra, $g$ and $h$ define an ideal in the ring $\mathds{C}[x,z,q]$. We want to find a basis of this ideal where one of the generators contains only $x$ and $z$. Such a basis is precisely given by the Groebner basis. Thus, we simply have to find the Groebner basis and extract the first basis element. Extra care is needed, since it can happen that the resulting polynomial is not irreducible.
 Since we know that $\tilde{P}_{\boldsymbol{w}}(x;z)$ is an irreducible polynomial, we are guaranteed that it appears as one of the factors in this basis element. By experimentation, we found that there is always only one irreducible factor that contains both $x$ and $z$, which we can hence identify with $\tilde{P}_{\boldsymbol{w}}(x;z)$. Finally, we normalise the polynomial such that the highest power of $z$ has unit coefficient, which follows from our definition \eqref{eq:def Ptilde poly}.
 
 This algorithm is implemented in the ancillary {\tt Mathematica} file and is very efficient at producing these polynomials. The function \texttt{Ptilde[w\_,x\_,z\_]} produces the polynomial and \texttt{Ptildenorm[w\_,x\_,z\_]} normalises it correctly. Finally, \texttt{P[w\_,x\_,z\_]} inserts all the prefactors that were discussed in Section~\ref{subsec:covering maps}. 
 
 \section{The function \texorpdfstring{$\boldsymbol{f(w)}$}{f(w)}} 
 \label{app:function fw}

The function $f(\boldsymbol{w})$, introduced in eq.~\eqref{eq:def P poly} reads
\begin{multline}
f(\boldsymbol{w})= \mathcal{S}(\boldsymbol{w}) \frac{\displaystyle\prod_{p=1}^{(|w_1-w_2|-|w_3-w_4|)/2} \binom{\max(w_3, w_4)+p-1}{\frac{|w_1-w_2|+|w_3-w_4|}{2}}}{\displaystyle\prod_{\ell=1}^{\left||w_1-w_2|-|w_3-w_4|\right|/2-1} \binom{\frac{|w_1-w_2|+|w_3-w_4|}{2}+\ell}{\ell}} \\
 \times \, \prod_{q=1}^{(|w_3-w_4|-|w_1-w_2|)/2}\binom{\max(w_1, w_2)+q-1}{\frac{|w_1-w_2|+|w_3-w_4|}{2}} \ , 
\end{multline}
where $\mathcal{S}(\boldsymbol{w})$ is a sign, depending on $\boldsymbol{w}$ mod $2$, 
\begin{subequations}
\begin{align} 
\boldsymbol{w} & \sim (0,0,0,0): & &(-1)^{\scaleto{\frac{s( (-w_1+w_2+w_3-w_4) (-w_1-w_2+w_3+w_4))}{4}
 +\frac{s(-w_1+w_2-w_3+w_4)}{2}+ \frac{s((-w_1-w_2+w_3+w_4)^2)}{8}+\frac{w_1+w_2+w_3+w_4}{4} }{10pt}}\ , \\
 \boldsymbol{w} & \sim (0,0,1,1): && (-1)^{ \scaleto{\frac{ s((-w_1+w_2-w_3+w_4)^2) }{8}+ \frac{ s( (-w_1+w_2-w_3+w_4) )}{2}+\frac{w_1-w_2+w_3-w_4}{4} }{10pt}}\ , \\
 \boldsymbol{w} &\sim (0,1,0,1): & & (-1)^{\scaleto{-\frac{s( (-w_1+w_2+w_3-w_4) (-w_1-w_2+w_3+w_4))}{4}+\frac{ s( (w_1+w_2-w_3-w_4)^2)}{8}+\frac{-w_1-w_2-w_3+w_4}{4} }{10pt}}\ , \\
 \boldsymbol{w} & \sim (0,1,1,0) : & &  (-1)^{\scaleto{ -\frac{ s( (-w_1-w_2+w_3+w_4)^2)}{8}+\frac{-w_1+w_2-w_3-w_4}{4} }{10pt}}\ ,\\
 \boldsymbol{w} & \sim (1,0,0,1): & &  (-1)^{\scaleto{-\frac{ s( (-w_1-w_2+w_3+w_4)^2)}{8}+\frac{w_1-w_2-w_3-w_4}{4} }{10pt}}  \ ,\\
 \boldsymbol{w} & \sim (1,0,1,0) : & &  (-1)^{\scaleto{ -\frac{s( (-w_1+w_2+w_3-w_4) (-w_1-w_2+w_3+w_4))}{4} +\frac{ s( (w_1+w_2-w_3-w_4)^2)}{8}+\frac{w_1 + w_2 - w_3 + w_4}{4} }{10pt} } \ ,	\\
 \boldsymbol{w} & \sim (1,1,0,0: & &  (-1)^{\scaleto{ \frac{ s((-w_1+w_2-w_3+w_4)^2) }{8} + \frac{s( (-w_1+w_2-w_3+w_4) )}{2}+\frac{-w_1+w_2+w_3-w_4}{4} }{10pt}} \ ,	\\
 \boldsymbol{w} & \sim (1,1,1,1): & & (-1)^{\scaleto{\frac{s( (-w_1+w_2+w_3-w_4) (-w_1-w_2+w_3+w_4)) }{4}+\frac{s(-w_1+w_2-w_3+w_4)}{2}+ \frac{s((-w_1-w_2+w_3+w_4)^2)}{8} -\frac{w_1+w_2-w_3-w_4}{4} }{10pt}}\ . 
\end{align}
\end{subequations}

\section{Identities for the polynomials \texorpdfstring{$\boldsymbol{P_{\boldsymbol{w}}(x;z)}$}{Pw(x,z)}} \label{app:P identities}
\label{app:Pw-identities}

There is a quadratic relation between the polynomials $P_{\boldsymbol{w}}(x;z)$ that ensures that combinations of $X_{ij}$'s behave like cross-ratios. It reads
\begin{multline}
\sqrt{z} P_{\boldsymbol{w}+\sum_{i=2,3}\varepsilon_i e_i}(x;z)P_{\boldsymbol{w}+\sum_{i=1,4}\varepsilon_i e_i}(x;z)-P_{\boldsymbol{w}+\sum_{i=1,2}\varepsilon_i e_i}(x;z)P_{\boldsymbol{w}+\sum_{i=3,4}\varepsilon_i e_i}(x;z)\\
+(-1)^{w_3w_4+w_2w_4+w_1w_3+w_1w_2}\sqrt{1-z}P_{\boldsymbol{w}+\sum_{i=1,3}\varepsilon_i e_i}(x;z)P_{\boldsymbol{w}+\sum_{i=2,4}\varepsilon_i e_i}(x;z)=0\ ,
\end{multline}
where $\varepsilon_i \in \{\pm 1\}$ and $e_i$ is the unit vector in the $i$-th direction.
This is equivalent to the identity 
\be 
X_{32}X_{41}-X_{21}X_{34}+X_{24}X_{31}=0
\ee
that is satisfied by the `generalised differences' $X_{ij}$, see \eqref{eq:def XI} for their definition.
Another equivalent form is
\begin{multline}
P_{\boldsymbol{w}+\sum_{i=2,3}\varepsilon_i e_i}(x;z)P_{\boldsymbol{w}+\sum_{i=1,4}\varepsilon_i e_i}(x;z)-\sqrt{z}P_{\boldsymbol{w}+\sum_{i=1,2}\varepsilon_i e_i}(x;z)P_{\boldsymbol{w}+\sum_{i=3,4}\varepsilon_i e_i}(x;z)\\
+(-1)^{w_3w_4+w_2w_4+w_1w_3+w_1w_2}\sqrt{1-z}P_{\boldsymbol{w}}(x;z)P_{\boldsymbol{w}+\sum_{i=1,2,3,4}\varepsilon_i e_i}(x;z)=0\ ,
\end{multline}
that makes the equality 
\be 
X_{32}X_{41}-z X_{21} X_{34}+X X_{1234}=0 \label{eq:Xij X X1234 relation}
\ee
manifest, see \eqref{eq:def XI} for the definition of these polynomials.

There is a second type of quadratic identity that involves no shifts in one of the four spectral flows. We can write it as
\be 
X_1 \big|_{\boldsymbol{w}\to \boldsymbol{w}+e_4}X_3 \big|_{\boldsymbol{w}\to \boldsymbol{w}-e_4}-X_1 \big|_{\boldsymbol{w}\to \boldsymbol{w}-e_4}X_3 \big|_{\boldsymbol{w}\to \boldsymbol{w}+e_4}=\pm X X_{13}\ . \label{eq:2nd quadratic identity}
\ee
We haven't tried to determine the sign in this identity since we won't need it.
This holds provided that 
\be 
\sum_{i=1}^4 w_i \in 2 \mathds{Z}\quad\text{and}\quad \sum_{i=1}^3 w_i \ge 1+2 \max_{i=1,2,3} w_i\ .
\ee
This identity is required to prove reflection symmetry of the spectrally flowed four-point function. There exist similar identities in the cases where the roles of the indices 1, 2 and 3 in the above expression have been permuted.
There is also a corresponding identity for the other parity, which takes the form
\be 
X \big|_{\boldsymbol{w}\to \boldsymbol{w}-e_4}X_{13} \big|_{\boldsymbol{w}\to \boldsymbol{w}+e_4}-X \big|_{\boldsymbol{w}\to \boldsymbol{w}+e_4}X_{13} \big|_{\boldsymbol{w}\to \boldsymbol{w}-e_4}= \pm X_1 X_{3}\ , \label{eq:3rd quadratic identity}
\ee
which holds provided that
\be 
\sum_{i=1}^4 w_i \in 2 \mathds{Z}+1\quad\text{and}\quad \sum_{i=1}^3 w_i \ge 1+2 \max_{i=1,2,3} w_i\ .
\ee

\section{Null vectors of \texorpdfstring{$\boldsymbol{\mathfrak{sl}(2,\mathds{R})_k}$}{sl(2,R)k}}
\label{app:null-vectors}

In this appendix we review the construction of $\mathfrak{sl}(2,\mathds{R})$ null vectors and present  examples that complement the discussion of Sections \ref{sec:null-vectors} and \ref{subsec:more constraints}. 

\subsection{Null vectors in the unflowed sector}

As already mentioned in the main text, $\mathfrak{sl}(2,\mathds{R})_k$ null vectors appear in the unflowed sector for \cite{Kac:1979fz}
\be 
j=j_{r,s}^+=\frac{1 + r +s(k-2)}{2} \ , \qquad r,s \in \mathds{Z}_{> 0}
\ee
or 
\be 
j=j_{r,s}^-=\frac{1 - r - s(k-2)}{2} \ , \qquad r \in \mathds{Z}_{> 0} \ , \quad s \in  \mathds{Z}_{\geq 0} \ . 
\ee
The null vector with spin $j_{r,s}^-$ can be written as \cite{Malikov1984}
\be
(J_0^+)^{r+st} (J_{-1}^-)^{r + (s-1)t} (J_0^+)^{r+(s-2)t} (J_{-1}^-)^{r + (s-3)t} \dots (J_0^+)^{r-st} \ket{j,j} \ , 
\label{null vector j minus}
\ee
where $t = -k + 2$. Similarly, for $j_{r,s}^+$ it reads
\be
(J_{-1}^-)^{r + (s-1)t} (J_0^+)^{r+(s-2)t} (J_{-1}^-)^{r + (s-3)t} (J_0^+)^{r+(s-4)t}  \dots (J_{-1}^-)^{r-(s-1)t} \ket{j,j} \ . 
\label{null vector j plus}
\ee
Note that for $s=0$, the general form \eqref{null vector j minus} reproduces the  two low-lying null vectors \eqref{null1} and \eqref{null2}. Similarly, for $r=s=1$ eq.~\eqref{null vector j plus} reproduces eq.~\eqref{null4}. Moreover, it is straightforward to check that also $\mathcal{N}=(J^-_{-1})^2\ket{j,j}$ with $j=j_{2,1}^+$ is null. 

Notice that the exponents in \eqref{null vector j minus} and \eqref{null vector j plus} need not be integer. Following \cite{Kac:1979fz,Bauer:1993jj}, let us review how by analytic continuation in the exponents eqs.~\eqref{null vector j minus} and \eqref{null vector j plus} give rise to null vectors. Let us consider the example 
\be 
j=j^-_{1,1} \ , \qquad \ket{\mathcal N} = (J_0^+)^{1+t}J_{-1}^-(J_0^+)^{1-t} \ket{j,j}
\label{null vector minus11}
\ee
where we have made use of \eqref{null vector j minus}. The expression in \eqref{null vector minus11} should be understood as follows. Expanding
\be 
e^{x J^+_0} J^-_{-1} e^{-x J_0^+} = J_{-1}^- - 2x J^3_{-1} + x^2 J^+_{-1}
\ee
in powers of $x$ one obtains
\be 
(J_0^+)^p J_{-1}^- = J_{-1}^-(J_0^+)^p - 2 p J_{-1}^3 (J_0^+)^{p-1} + p (p-1) J^+_{-1} (J_0^+)^{p-2} \ . 
\ee
Setting $p=t+1$ and analytically continuing in $t$, we find 
\begin{align}
& (J_0^+)^{1+t}J_{-1}^-(J_0^+)^{1-t} \nonumber\\
 & \quad =  \Bigl(J_{-1}^-(J_0^+)^{t+1} - 2 (t+1) J_{-1}^3 (J_0^+)^{t} + t (t+1) J^+_{-1} (J_0^+)^{t-1} \Bigr) (J_0^+)^{1-t} \\
 & \quad = J^{-}_{-1} (J_0^+)^2 + 2(k-3) J^3_{-1}J^+_0 + (k-2)(k-3) J^+_{-1} \ , 
\end{align}
which, as expected, agrees with \eqref{null3}. Let us consider an additional example,
\be 
j = j_{2,1}^- \ , \qquad \ket{\mathcal{N} }= (J_0^+)^{2+t} (J^-_{-1})^2 (J_0^+)^{2-t}\ket{j,j}
\ee
With the help of {\tt Mathematica} and making use of the \textit{Virasoro} package
developed by Matthew Headrick \cite{Virasoro.package} one finds
\begin{align}
(J^+_0)^p (J_{-1}^-)^2 = & \  (J_{-1}^-)^2 (J^+_0)^p + 2 p J^-_{-2} (J_0^+)^{p-1}-4p J^-_{-1} J^3_{-1} (J_0^+)^{p-1} \nonumber \\
&  -2p(p-1) J^3_{-2} (J_0^+)^{p-2} + 2p(p-1) J^-_{-1} J^+_{-1} (J_0^+)^{p-2} \nonumber \\ 
& + 4p(p-1) J^3_{-1} J^3_{-1} (J_0^+)^{p-2} + 2p(p-1)(p-2) J^+_{-2} (J_0^+)^{p-3}  \nonumber \\ 
& -4p(p-1)(p-2) J^3_{-1} J^+_{-1} (J_0^+)^{p-3}  \nonumber \\ 
& + p (p-1)(p-2)(p-3) J^+_{-1} J^+_{-1} (J_0^+) ^{p-4} \ . 
\end{align}
Following the same strategy of the previous example, for $p=t+2$ we obtain 
\begin{align}
\ket{\mathcal N} = & \ (J_0^+)^{2+t} (J^-_{-1})^2 (J_0^+)^{2-t} \ket{j,j}\\
= & \Biggl( (J_{-1}^-)^2 (J^+_0)^4 + 2 (t+2) J^-_{-2} (J_0^+)^3\nonumber \\ 
& \quad -4(t-2) J^-_{-1} J^3_{-1} (J_0^+)^3  -2(t+1)(t+2) J^3_{-2} (J_0^+)^2 \nonumber \\ 
& \quad \quad + 2(t+1)(t+2) J^-_{-1} J^+_{-1} (J_0^+)^2  + 4(t+1)(t+2) J^3_{-1} J^3_{-1} (J_0^+)^2  \nonumber \\ 
& \quad \quad \quad  + 2t(t+1)(t+2) J^+_{-2} J_0^+  -4t(t+1)(t+2) J^3_{-1} J^+_{-1} J_0^+ \nonumber \\
& \hspace{175pt} + (t-1)t(t+1)(t+2) J^+_{-1} J^+_{-1} \Biggr) \ket{j,j}  \ . 
\label{null vector j minus21}
\end{align}
Once more making use of {\tt Mathematica} and of the \textit{Virasoro} package \cite{Virasoro.package}, one can check that indeed the vector in eq.\ \eqref{null vector j minus21} is null and annihilated by $J^-_0$ for $j= j_{2,1}^-$ and $t= -k+2$. 

\medskip 

From eqs.~\eqref{null vector j minus} and \eqref{null vector j plus} it follows that null vectors with spin $j=j^\pm_{r,s}$ appear at level $r s$. The degree of the associated differential equation coincides with the number of currents. It is then given by 
\begin{align}
\text{ord}(j^-_{r,s})& = r \sum_{n=-s}^s 1 = r(2s + 1) \ , \\ 
\text{ord}(j^+_{r,s})& = r \sum_{n=-(s-1)}^{s-1} 1 = r(2s - 1) \ .
\end{align}

\subsection{Null vector differential equations}

Null vectors give rise to differential equations for correlators with one degenerate field. Let us  present a few examples that complement the discussion of Sections \ref{sec:null-vectors} and \ref{subsec:more constraints}. We discuss null vector equations both in the unflowed \cite{Teschner:1997ft} and flowed sector and investigate their relation. 

\subsubsection{The unflowed sector}

For $j_4=j_{2,0}^-=-\frac{1}{2}$, the differential equation associated to the null vector \eqref{null2} is simply 
\be 
\partial_{x_4}^2 \left\langle \prod_{i=1}^3 V^0_{j_i}(x_i;z_i) \, V^0_{j_4}(x_4;z_4) \right\rangle =0 \ . 
\label{eq:unflowed-null-eq j=-1/2}
\ee
Slightly more involved is the derivation of the null equation for \eqref{null3}. Contour deformation arguments imply for $j_4 = j_{1,1}^-=1-\frac{k}{2}$
\begin{multline}
\sum_{i=1}^3\frac{1}{z_4-z_i} \Biggl[ (x_i-x_4)^2 \, \partial_{x_4}^2 \, \partial_{x_i} + 2 (x_i-x_4 ) \, j_i \, \partial_{x_4}^2 + 2(k-3)(x_i-x_4) \, \partial_{x_4} \, \partial_{x_i} \\
+2(k-3) \, j_i \,  \partial_{x_4} + (k-3)(k-2) \, \partial_{x_i} \Biggr] \left\langle \prod_{\ell=1}^4 V^0_{j_\ell}(x_\ell;z_\ell) \right\rangle =0 \ . 
\label{null equation j=1-k/2}
\end{multline}
Making use of global Ward identities and choosing the insertion points as 
\be 
z_1=x_1=0 \ , \qquad z_2=x_2=1 \ , \qquad  z_3=x_3= \infty \ , 
\ee
eq.~\eqref{null equation j=1-k/2} can be rewritten as 
\begin{align}
& \hspace{-5pt} \Biggl[ -2 \left(x_4-1\right) x_4 \left(x_4-z_4\right) \partial_{x_4}^3+ \Bigl(x_4^2 (-2 j_1-2 j_2+2 j_3+5 k-18) + \nonumber \\
& \hspace{0pt} -2 x_4( 2 j_3 z_4+k z_4-6 z_4-2 j_1+2k-6)+z_4 (-2 j_1+2 j_2+2 j_3+k-6)\Bigr)\partial^2_{x_4} \nonumber \\
& \hspace{0pt} -2 (k-3)\Bigl(2 x_4(-j_1-j_2+j_3+k-3)+2 j_1 -2 j_3 z_4-k+2 z_4+2\Bigr) \partial_{x_4} \nonumber \\
& \hspace{65pt} +(k^2-5 k+6) (-2 j_1-2 j_2+2 j_3+k-2) \Biggr] \left\langle \prod_{\ell=1}^4 V^0_{j_\ell}(x_\ell;z_\ell) \right\rangle = 0 \ . 
\label{unflowed null equation j=1-k/2}
\end{align}  
 
\subsubsection{The flowed sector}

Let us derive the null vector equation for $j_4 = -\frac{1}{2}$ and $\sum_i w_i \in 2 \mathds{Z}$ , $w_4>0$. The null field is 
\be
(h-\tfrac{kw}{2}-\tfrac{1}{2})(h-\tfrac{kw}{2}+\tfrac{1}{2})V_{-\frac{1}{2},h}^w(x;z) \longrightarrow y_4^2 \partial_{y_4}^2 V_{-\frac{1}{2}}^w(x;y;z)
\ee
and in terms of the $y$-transform the differential equation reads, see eq.~\eqref{eq:main conjecture even},
\be 
\partial_{y_4}^2 \Bigl( X_{34} \, F(c,z) \Bigr) = 0 \ , \qquad \text{with} \qquad c \equiv \frac{X_{23}X_{14}}{X_{12}X_{34}} \ . 
\ee
Distributing derivatives and using the chain rule it follows
\be 
\Bigl(X_{34} \,  \partial^2_{y_4} c + 2 (\partial_{y_4} X_{34}) \, \partial_{y_4}c \Bigr) \, \partial_c F(c,z) + X_{34} \, (\partial_{y_4}c)^2 \, \partial_c^2 F(c,z) = 0 \ . 
\label{flowed-null-eq-j=-1/2}
\ee
Using that $c=\frac{X_{23}X_{14}}{X_{12}X_{34}}$, the first term in the parenthesis vanishes. The differential equation obeyed by $F(c,z)$ then simply becomes 
\be 
\partial_c^2 F(c,z)= 0 \ , 
\ee
which coincides with the unflowed null vector equation \eqref{eq:unflowed-null-eq j=-1/2}.

\medskip

We now consider a more involved example, $j_4 = 1-\tfrac{k}{2}$ with $\sum_i w_i \in 2 \mathds{Z}$ and $w_4>0$. The null field is, see eq.~\eqref{null4m}
\begin{multline}
(k-2 h + kw-2) (k-2 h + kw) [J^+_{-2}V_{j,h-1}^w](x;z) \\
+2 (k-2h + k w-2) (k+2h - kw-2) [J^3_{-1}V_{j,h}^w](x;z)\\
+(k+2h - k w-2) (k+2h - k w) [J_{0}^- V_{j,h+1}^w](x;z)  \ .
 \label{null-vector-j4=1-k/2}
\end{multline}
Given a polynomial $p(h)$, from \eqref{eq:definition y basis} it follows 
\be 
p(h) V_{j,h,\bar{h}}^w(x;z)= \int \mathrm{d}^2 y \, y^{\frac{kw}{2}+j-h-1} \, \bar{y}^{\frac{kw}{2}+j-\bar{h}-1} \, \mathcal{D}_y^{p} V_j^w(x;y;z) \ ,
\label{eq:diff-operator-y-basis}
\ee
where $\mathcal{D}_y^{p}$ is a differential operator in $y$. In particular, we have 
\begin{align}
& V_{j,h-1,\bar{h}}^w(x;z)  \longrightarrow y \ , \qquad V_{j,h,\bar{h}}^w(x;z)  \longrightarrow 1 \ , \qquad V_{j,h+1,\bar{h}}^w(x;z)  \longrightarrow y^{-1} \ , \\
& h \, V_{j,h-1,\bar{h}}^w(x;z) \longrightarrow (j + \tfrac{k w}{2} + 1) y + y^2 \partial_y \ , \\
& h \, V_{j,h,\bar{h}}^w(x;z) \longrightarrow 	j + \tfrac{k w}{2} + y \partial_y \ , \\
& h \, V_{j,h+1,\bar{h}}^w(x;z) \longrightarrow (j-1 + \tfrac{k w}{2}) y^{-1} + \partial_y \ , \\
& h^2 V_{j,h-1,\bar{h}}^w(x;z) \longrightarrow (j^2  +2 j +j k w+ \tfrac{k^2 w^2}{4} + k w  +1  ) y \nonumber \\ 
& \hspace{100pt}+ (3+ 2 j + k w) y^2 \partial_y + y^3 \partial_y^2 \ , \\
& h^2 V_{j,h,\bar{h}}^w(x;z) \longrightarrow (j^2+j k w+\tfrac{k^2 w^2}{4} )+ (2 j+k w+1) \partial_y + y^2 \partial_y^2 \ , \\
& h^2 V_{j,h+1,\bar{h}}^w(x;z) \longrightarrow \tfrac{1}{4} (2 j +k w-2)^2 y^{-1}+ (2 j+k w-1)\partial_y + y \partial_y^2 \ , 
\label{eq:h2Vhp1}
\end{align}
where the arrows are a shorthand for identities of the form \eqref{eq:diff-operator-y-basis}. Making use of equations \eqref{null-vector-j4=1-k/2} -- \eqref{eq:h2Vhp1} we obtain the differential equation
\begin{multline} 
4 \,  y_4 \Bigl( y_4^2 \partial_{y_4}^2 -2 (k-3) y_4 \partial_{y_4} + k^2-5 k+6 \Bigr) \left\langle \prod_{\ell=1}^3 V^{w_\ell}_{J_\ell}(x_{\ell} ;y_{\ell} ; z_{\ell}) \Bigl[J^+_{-2}V^{w_4}_{1-\frac{k}{2}}\Bigr](x_4 ;y_4 ; z_4) \right\rangle  \\
 + 8 \, y_4 \Bigl((k-3) \partial_{y_4} -y_4 \partial_{y_4}^2\Bigr) \left\langle \prod_{\ell=1}^3 V^{w_\ell}_{J_\ell}(x_{\ell} ;y_{\ell} ; z_{\ell}) \Bigl[J^3_{-1}V^{w_4}_{1-\frac{k}{2}}\Bigr](x_4 ;y_4 ; z_4) \right\rangle \\
 + 4 \, y_4 \, \partial_{y_4}^2 \left\langle \prod_{\ell=1}^3 V^{w_\ell}_{J_\ell}(x_{\ell} ;y_{\ell} ; z_{\ell}) \Bigl[J^-_{0}V^{w_4}_{1-\frac{k}{2}}\Bigr](x_4 ;y_4 ; z_4) \right\rangle = 0 \ .  
 \label{flowed third order null equation}
\end{multline}
We have checked extensively that upon making use of the recursion relations of \cite{Eberhardt:2019ywk}, \eqref{flowed third order null equation} reduces to a differential equation for $F(c,z)$, exactly reproducing \eqref{unflowed null equation j=1-k/2} with $F(c,z)$ in place of the unflowed correlator. The details can be found in the ancillary {\tt Mathematica} notebook. 

\section{An identity for the unflowed correlator}
\label{app:identity}

In this appendix we derive the identity
\begin{multline}
\mathcal N(j_1) \left\langle V^0_{\frac{k}{2}-j_1}(0;0) V^0_{j_2}(1;1) V^0_{j_3}(\infty;\infty) V^0_{j_4}(x;z) \right\rangle \\
= \mathcal N(j_3) |x|^{-4j_4} |z|^{2j_4} \left\langle V^0_{j_1}(0;0) V^0_{j_2}(1;1) V^0_{\frac{k}{2}-j_3}(\infty;\infty) V^0_{j_4}\Bigl(\frac{z}{x};z \Bigr) \right\rangle \ . 
\label{identity-app}
\end{multline}
Eq.~\eqref{identity-app} has been shown in \cite{Parnachev:2001gw}.  The derivation we present here is alternative and some of the steps involved will be useful throughout the text. Let us start by considering the five-point function
\be
\left\langle V^0_{j_1}(0;0) V^0_{j_2}(1;1) V^0_{j_3}(\infty;\infty) V^0_{j_4}(x_4;z_4) V^0_{\frac{k}{2}}(x_5;z_5) \right\rangle \ . 
\label{5ptf}
\ee
Making use of eq.~\eqref{sugawara}, two independent KZ differential equations can be derived along the lines of \cite{Teschner:1999ug}. Moreover, since the spin $j_5=j^+_{1,1} =\frac{k}{2}$ is degenerate, the correlator \eqref{5ptf} also obeys one additional null vector equation. By solving one of the two KZ equations and the null vector equation --- see the ancillary {\tt Mathematica} notebook for the explicit computation --- one finds 
\begin{multline}
\left\langle V^0_{j_1}(0;0) V^0_{j_2}(1;1) V^0_{j_3}(\infty;\infty) V^0_{j_4}(x_4;z_4) V^0_{\frac{k}{2}}(x_5;z_5) \right\rangle \\
= |1-x_5|^{2(-\frac{k}{2}+j_1-j_2+j_3+j_4)} |x_5|^{2(-\frac{k}{2}-j_1+j_2+j_3+j_4)} |x_4-x_5|^{-4j_4} |x_5-z_5|^{2(\frac{k}{2}-j_1-j_2-j_3-j_4)} \\
\times |1-z_5|^{2j_2} |z_5|^{2j_1} |z_4-z_5|^{2j_4} f\left(\frac{(1-x_5)(x_5 z_4 - x_4 z_5)}{(x_4-x_5)(x_5-z_5)},z_4\right) \ , 
\label{KZ-null-sol}
\end{multline}
where $f(x,z)$ is an unknown function of two variables. 

Let us now consider the $z_5 \to 0$ limit of \eqref{5ptf} and fuse $V^0_{\frac{k}{2}}(x_5;z_5)$ with $V^0_{j_1}(0;0)$. We find 
\begin{multline}
\left\langle V^0_{j_1}(0;0) V^0_{j_2}(1;1) V^0_{j_3}(\infty;\infty) V^0_{j_4}(x_4;z_4) V^0_{\frac{k}{2}}(x_5;z_5) \right\rangle \\
\sim |z_5|^{2(\Delta(\frac{k}{2}-j_1)-\Delta(j_1)-\Delta(\frac{k}{2}))} \int \text{d}j \int \text{d}^2 x \, |x_5|^{2(1-j_1-\frac{k}{2}-j)} |x|^{2(\frac{k}{2}-1-j_1+j)} |x_5-x|^{2(j_1-\frac{k}{2}-1+j)}  \\
\times \, D(1-j, j_1, \tfrac{k}{2}) \left\langle V^0_{j}(x;0) V^0_{j_2}(1;1) V^0_{j_3}(\infty;\infty) V^0_{j_4}(x_4;z_4) \right\rangle \ , 
\label{z50fusion}
\end{multline}
where we used the explicit form of the OPEs given in \cite{Teschner:1997ft} and $\Delta(j)$ is the conformal dimension on the worldsheet,  
\be 
\Delta(j) \equiv -\frac{j(j-1)}{k-2} \ . 
\ee
It follows from \cite[eqs.~(B.7) and (B.8c)]{Dei:2021xgh} that
\be
D(1-j,j_1, \tfrac{k}{2}+\epsilon) = -\frac{\epsilon \, B(j)}{\pi B(0)}\, \delta(j-\tfrac{k}{2}+j_1) \ , 
\label{onechannel}
\ee
where following \cite{Teschner:1999ug} we regularised the singularity by the replacement $j_5=\frac{k}{2} \to \frac{k}{2}+ \epsilon$. Adopting the same regularisation in \eqref{z50fusion} and making use of \eqref{onechannel} we obtain for $z_5 \to 0$
\begin{multline}
\left\langle V^0_{j_1}(0;0) V^0_{j_2}(1;1) V^0_{j_3}(\infty;\infty) V^0_{j_4}(x_4;z_4) V^0_{\frac{k}{2}}(x_5;z_5) \right\rangle \\
\sim  -\frac{\epsilon \, B(j)}{\pi B(0)} |z_5|^{2j_1} |x_5|^{2(-k+1)} \int \text{d}^2 x |x|^{2(-2j_1+k-1)}|x_5-x|^{2(-1-\epsilon)} \\
\left\langle V^0_{\frac{k}{2}-j_1}(x;0) V^0_{j_2}(1;1) V^0_{j_3}(\infty;\infty) V^0_{j_4}(x_4;z_4) \right\rangle \ . 
\end{multline}
We regulate the divergence in the integral as in \cite{Teschner:1999ug}, and formally write
\be
|x_5-x|^{2(-1-\epsilon)} = -\frac{\pi \delta^2(x_5-x)}{\epsilon} \ . 
\ee
Hence, for $z_5 \to 0$
\begin{multline}
\left\langle V^0_{j_1}(0;0) V^0_{j_2}(1;1) V^0_{j_3}(\infty;\infty) V^0_{j_4}(x_4;z_4) V^0_{\frac{k}{2}}(x_5;z_5) \right\rangle \\
\sim \frac{B(j_1)}{B(0)}  |z_5|^{2j_1} |x_5|^{-4j_1} \left\langle V^0_{\frac{k}{2}-j_1}(x_5;0) V^0_{j_2}(1;1) V^0_{j_3}(\infty;\infty) V^0_{j_4}(x_4;z_4) \right\rangle \ . 
\end{multline}
Finally, comparing with \eqref{KZ-null-sol} and setting $x_5 = 0$, 
\be 
\left\langle V^0_{\frac{k}{2}-j_1}(0;0) V^0_{j_2}(1;1) V^0_{j_3}(\infty;\infty) V^0_{j_4}(x_4;z_4) \right\rangle \\
= \frac{B(j_1)}{B(0)} |x_4|^{-4j_4} |z_4|^{2j_4} f\left(\frac{z_4}{x_4},z_4\right) \ . 
\label{fusion-0}
\ee

By similar techniques one can consider the fusion of $V^0_{\frac{k}{2}}(x_5;z_5)$ with $V_{j_3}^0(\infty;\infty)$ and obtain
\be
f(x_4,z_4) = \frac{B (j_3)}{B(0)} \left\langle V^0_{j_1}(0;0) V^0_{j_2}(1;1) V^0_{\frac{k}{2}-j_3}(\infty;\infty) V^0_{j_4}(x_4;z_4) \right\rangle
\label{fusion-inf}
\ee
Injecting \eqref{fusion-inf} into \eqref{fusion-0} and noticing that 
\be 
\frac{B\left(j_1\right)}{B\left(j_3\right)} = \frac{\mathcal{N}(j_1)}{\mathcal{N}(j_3)}
\ee
one recovers eq.~\eqref{identity-app}.

\bibliographystyle{JHEP}
\bibliography{bib}
\end{document}